\documentclass[11pt, a4paper]{article}
\usepackage{amsmath}
\usepackage{amssymb}

\newcommand{\be}{\begin{equation}}
\newcommand{\ee}{\end{equation}}
\newcommand{\bea}{\begin{eqnarray}}
\newcommand{\eea}{\end{eqnarray}}
\newcommand{\ba}{\begin{array}}
\newcommand{\ea}{\end{array}}

\begin{document}
\begin{center}
\baselineskip 20pt
{\Large\bf
Goldstone Modes  in Renormalizable Supersymmetric SO(10) Model
}
\vspace{1cm}

{\large
Xian-Zheng Bai \footnote{ E-mail: 15012100098@pku.edu.cn}
and
Da-Xin Zhang\footnote{ E-mail: dxzhang@pku.edu.cn}
}
\vspace{.5cm}

{\baselineskip 20pt \it
School of Physics and State Key Laboratory of Nuclear Physics and Technology, \\
Peking University, Beijing 100871, China}

\vspace{.5cm}

\vspace{1.5cm} {\bf Abstract}
\end{center}

We solve the Goldstone modes in the renormalizable SUSY SO(10) model with general couplings.
The Goldstones are expressed by the Vacuum Expectation Values and the  Clebsch-Gordan coefficients
of relevant symmetries without explicit dependence on the parameters of the model.

\section{Introduction}
Grand Unified Theories (GUTs) are very important candidates for the physics beyond
the Standard Model (SM).
The Pati-Salam  $G_{422} = SU(4)_C \otimes SU(2)_L \otimes SU(2)_R$ \cite{PS},
the Georgi-Glashow $SU(5)$\cite{su5}, the flipped $SU(5)$ of
$G_{51}=(SU(5) \otimes U(1))^{Flipped}$  \cite{flip}, the SO(10) \cite{so10a,so10b}
and the $E_6$\cite{E6} models have all been studied extensively in the literature.
Except $SU(5)$, the GUT symmetries of these models are larger in ranks than the SM symmetry
 $G_{321} = SU(3)_C \otimes SU(2)_L \otimes U(1)_Y$.
Consequently, they have several different routines to break into the SM symmetry,
since the SM symmetry $G_{321}$ is not the maximal subgroup of the GUT symmetries.

The supersymmetric (SUSY) GUT models are even more important in
 realizing gauge coupling unification\cite{sen1,sen2,luo1,luo2,luo3,luo4,unif1,unif2,unif3}.
Again, except in the $SU(5)$ models, these GUT symmetries have different routines
in the symmetry breaking chains.

Spontaneously Symmetry Breaking (SSB) generates the massless Goldstone bosons corresponding to the
broken generators\cite{gold1,gold2,gold3}. In gauge theories, these Goldstone bosons act as
the longitudal components of the gauge bosons.
In model buildings, we need to check the spectra to verify the existence of
these Goldstone bosons.
When more than one Higgs multiplets contribute to the SSB, we need to check
that there are massless eigenstates in the mass squared matrices with correct representations.

In SUSY models, there are easier ways  to get the Goldtone bosons.
The Goldstinos, which are the fermionic partners of the Goldstone bosons,
 are also massless
which mix with gauginos, the SUSY partners of the gauge bosons.
Then, we can get the Goldstinos by solving the massless eigenstates of the mass matrices.
Being the SUSY partners, the Goldstone bosons are determined accordingly.

Usually in building GUT models,
the Higgs sectors relevant to SSB are complicated. Hereafter we will focus on
SUSY SO(10)\cite{kuo,moha,he,lee,sato,np597,so10c,bajc,fuku1,fuku2,fuku,np711,np757,malinsky,np857,np882,nath2015,czy}.
Even in the minimal renormalizble SUSY SO(10)
model \cite{bajc}, Higgs in $210+126+\overline{126}$ are needed to break SO(10)
into $G_{321}$. There are Goldstinos in $(1,1,0),(1,1,1),(3,1,\frac{2}{3}),
(3,2, -\frac{5}{6}),(3,2,\frac{1}{6})$ and the conjugates under $G_{321}$.
None of  these representations has only one field in the spectrum.
To verify the Goldstone modes,
usually people need to calculate the eigenvalues of the mass matrices
for the Goldstinos\cite{bajc,malinsky}.
In more general models with also $45+54$ in the SSB of the GUT symmetry,
the Goldstino check is done numerically\cite{sato,fuku}.

In this work, we will study the Goldstone modes within SUSY.
In \cite{rzx} it has been realized that for the SSB of $U(1)$ symmetries in SUSY models,
in the Goldstone mode a component is proportional to the charge and to the Vacuum Expectation Value (VEV).
Furthermore, it has been shown
in the Non-Abelian cases \cite{su2cgc},
that in any Goldstone mode, the component from a representation is proportional to the VEV from
the same representation and the CGCs determine the remainder dependence.
Examples on the simplest $SU(2)$ SSB are given in \cite{su2cgc}.
Here we will show test explicitly in the SUSY SO(10) model.
We will give the Goldstone modes for the SSB of SO(10) into $G_{321}$ in the general renormalizable model.

In Section 2 we will give a general discussion on the method to solve the Goldstone modes.
Then, in Section 3, we associate in $SO(10)$ the Goldstones and  the
corresponding broken symmetries.
In Section 4 and 5,  Goldstones associated with the SSB of $G_{422}$ and $SU(5)$ are solved, respectively,
and relevant identities among the CGCs are examined.
In Section 6, Goldstones associated with the SSB of the flipped $SU(5)$ are solved.
The relevant CGCs,  which are not available in the
literature, are presented and the identities are examined. The mass matrix for the Goldstones is
given in the Appendix.
Finally we summarize in Section 7.

\section{General SSB in SUSY Models}

Following \cite{su2cgc},
we consider the SSB of $G_1 \rightarrow G_2$ in SUSY models.
For models with only real fields, the general Higgs superpotential can be written as
\begin{equation}
W = \frac{1}{2} \sum_{I,i} m_I (I_i)^2
+ \sum_{IJK,ijk} {\lambda}^{IJK} I_i J_j K_k,\label{real1}
\end{equation}
where $I,J,K$ denotes the superfields, $i,j,k$ denote the representations of group $G_1$.
Under $G_2$, the couplings in (\ref{real1}) are of the forms
\begin{equation}
{\lambda}^{IJK} I_{i_a} J_{j_b} K_{k_c} C^{IJK}_{i_a j_b k_c},\label{real2}
\end{equation}
where $a,b,c$ are the representations under $G_2$, and $C^{IJK}_{i_a j_b k_c}$ is the Clebsch-Gordan coefficient (CGC).
A singlet of $G_2$, denoted by $I_{i_1}$ {\it etc.},
can have a VEV  $\hat{I}_{i_1}$ in the SSB.
SUSY requires the F-flatness conditions for the singlets,
\begin{equation}
0\equiv F_{I_{i_1}} = \langle \frac{\partial W}{\partial I_{i_1}} \rangle
= m_I \hat I_{i_1} + \sum_{JK,jk} \lambda^{IJK} \hat J_{j_1} \hat K_{k_1} C^{IJK}_{{i_1}{j_1}{k_1}}.
\label{realF}
\end{equation}
The Goldstinos
in the SSB will be denoted by $\alpha,\overline{\alpha}$ under $G_2$.
A mass matrix element for the Goldstinos is
\begin{equation}
M^{IJ}_{{i_\alpha}{j_{\bar{\alpha}}}}
= \langle \frac{\partial^2 W}{\partial I_{i_\alpha} J_{j_{\bar{\alpha}}}} \rangle
= m_I \delta_{IJ} \delta_{ij}
+ \sum_{K,k} \lambda^{IJK} \hat K_{k_1} C^{IJK}_{{i_\alpha}{j_{\bar{\alpha}}}{k_1}}.\nonumber
\end{equation}
For $G^I_{i_{\bar{\alpha}}}$ denotes the element of the Goldstino corresponding to $\bar{\alpha}$,
the zero-eigenvalue equation is
\begin{equation}
0\equiv \sum_{J,j} M^{IJ}_{{i_\alpha}{j_{\bar{\alpha}}}} G^J_{j_{\bar{\alpha}}}
= m_I G^I_{i_{\bar{\alpha}}} + \sum_{JK,jk} \lambda^{IJK} G^J_{j_{\bar{\alpha}}}
\hat K_{k_1} C^{IJK}_{{i_\alpha}{j_{\bar{\alpha}}}{k_1}}.  \label{realM}
\end{equation}
Eliminating $M_I$ in (\ref{realF},\ref{realM}) gives
\begin{equation}
0 = G^I_{i_{\bar{\alpha}}} \sum\limits_{JK,jk} \lambda^{IJK} \hat J_{j_1} \hat K_{k_1} C^{IJK}_{{i_1}{j_1}{k_1}} -  \hat I_{i_1} \sum\limits_{JK,jk} \lambda^{IJK} G^J_{j_{\bar{\alpha}}}
\hat K_{k_1} C^{IJK}_{{i_\alpha}{j_{\bar{\alpha}}}{k_1}}.\label{e6}
\end{equation}
It follows that if a representations of $G_1$ does not contain a $G_2$ singlet,
even if it may contain representations of the Goldstinos,
it does not contribute to the Goldstino modes since the $M_I$ term in (\ref{realM}) cannot be eliminated unless
multiplied by zero.

The superpotential parameters can be arbitrary so that we can focus on a specified coupling
$\lambda^{IJK}$, and the summation over different $\lambda$s is unnecessary.
Furthermore, denoting
\begin{equation}
T^{Ii}_{\bar{\alpha}} = \frac{G^I_{i_{\bar{\alpha}}}}{\hat I_{i_1}}, ~etc., \label{T}
\end{equation}
(\ref{e6}) becomes
\begin{equation}
0 = T^{Ii}_{\bar{\alpha}} \lambda^{IJK} \hat J_{j_1} \hat K_{k_1} C^{IJK}_{{i_1}{j_1}{k_1}}
- T^{Jj}_{\bar{\alpha}} \lambda^{IJK} \hat J_{j_1} \hat K_{k_1} C^{IJK}_{{i_\alpha}{j_{\bar{\alpha}}}{k_1}}
- T^{Kk}_{\bar{\alpha}} \lambda^{IJK} \hat J_{j_1} \hat K_{k_1} C^{IJK}_{{i_\alpha}{j_1}{k_{\bar{\alpha}}}}. \nonumber
\end{equation}
The nonzero $\lambda^{IJK}\hat J_{j_1} \hat K_{k_1}$ can be eliminated and we can
reiterate the same operation for $J,K$, then
\begin{equation}
0 = \left( \begin{array}{ccc}
- C^{IJK}_{{i_1}{j_1}{k_1}} &  C^{IJK}_{{i_\alpha}{j_{\bar{\alpha}}}{k_1}} &  C^{IKJ}_{{i_\alpha}{k_{\bar{\alpha}}}{j_1}} \\ \\
 C^{JIK}_{{j_\alpha}{i_{\bar{\alpha}}}{k_1}} & - C^{JIK}_{{j_1}{i_1}{k_1}} &
C^{JKI}_{{j_\alpha}{k_{\bar{\alpha}}}{i_1}} \\ \\
 C^{KIJ}_{{k_\alpha}{i_{\bar{\alpha}}}{j_1}} &  C^{KJI}_{{k_\alpha}{j_{\bar{\alpha}}}{i_1}} &
-C^{KJI}_{{k_1}{j_1}{i_1}}
\end{array}
\right)
\left(
\begin{array}{c}
T^{Ii}_{\bar{\alpha}} \\ \\ T^{Jj}_{\bar{\alpha}} \\ \\ T^{Kk}_{\bar{\alpha}}
\end{array}
\right). \label{19}
\end{equation}
Similar result is given for the Goldstino in $\alpha$.
({\ref{19}) is followed by an identity that the determinant of the square matrix in ({\ref{19}) is zero,
which must hold for any SSB.

Some special cases need to be clarified now.

(1) If $k$ does not contain $\alpha, {\bar{\alpha}}$, (\ref{19}) is simplified into
\begin{equation}
0 = \left( \begin{array}{cc}
- C^{IJK}_{{i_1}{j_1}{k_1}} &  C^{IJK}_{{i_\alpha}{j_{\bar{\alpha}}}{k_1}}  \\ \\
 C^{JIK}_{{j_\alpha}{i_{\bar{\alpha}}}{k_1}} & - C^{JIK}_{{j_1}{i_1}{k_1}}
\end{array}
\right)
\left(
\begin{array}{c}
T^{Ii}_{\bar{\alpha}} \\ \\ T^{Jj}_{\bar{\alpha}}
\end{array}
\right), \label{case1}
\end{equation}
which gives an identity that the determinant of the square matrix is zero, and  a simple ratio between the two components.
Taking all different couplings $\lambda$s setup all relations among the Goldstino components.
This determines the Goldstino content up to an overall normalization.

(2) For $J = K$ and $j = k$, we have
\begin{equation}
0 = \left( \begin{array}{cc}
- C^{IJJ}_{{i_1}{j_1}{j_1}} &  2 C^{IJJ}_{{i_\alpha}{j_{\bar{\alpha}}}{j_1}}  \\ \\
C^{IJJ}_{{i_{\bar{\alpha}}}{j_{\alpha}}{j_1}} & C^{IJJ}_{{i_1}{j_\alpha}{j_{\bar{\alpha}}}} - C^{IJJ}_{{i_1}{j_1}{j_1}}
\end{array}
\right)
\left(
\begin{array}{c}
T^{Ii}_{\bar{\alpha}} \\ \\ T^{Ij}_{\bar{\alpha}}
\end{array}
\right).\label{case2}
\end{equation}

(3) For $I = J = K$ and $i = j = k$, the result is
\begin{equation}
C^{III}_{{i_1}{i_1}{i_1}} =2 C^{III}_{i_{\alpha} i_1 i_{{\bar{\alpha}}}},\label{case3}
\end{equation}
without information on $T$.

Generalization to the models
with complex fields is straightforward with the general results given in \cite{su2cgc}.
We have also identities among the CGCs and equations for the Goldstinos involving the CGCs.
Two types of special superpotentials might be relevant in the $SO(10)$ study.

(4) For $I,\overline{I},K$ contain Goldstones and $K$ is real, if
\be
W=m_I I\overline{I}+\frac{1}{2}m_K^2+\lambda I^2K+\bar{\lambda}\overline{I}^2K,\nonumber
\ee
which gives
\be
0=\left(
\ba{ccc}
\ba{c}
-C^{IIK}_{{i_1}{i_1}{k_1}} \\ \\
XC^{\bar{I}\bar{I} K}_{\bar{i}_{\bar{\alpha}} {\bar{i}_1} {k_\alpha} }\\ \\
C^{\bar{I}\bar{I} K}_{\bar{i}_{\bar{\alpha}}{\bar{i}_\alpha}{k_1}}
\ea
\ba{c}
C^{IIK}_{{i_{\alpha}}{i_1}{k_{\bar{\alpha}}}}\\ \\
-C^{IIK}_{{i_1}{i_1}{k_1}} \\ \\
C^{\bar{I}\bar{I} K}_{\bar{i}_{\bar{\alpha}}{\bar{i}_1}{k_\alpha}}
\ea
\ba{c}
C^{IIK}_{{i_{\alpha}}{i_{\bar{\alpha}}}{k_1}}\\ \\
C^{IIK}_{{i_{\bar{\alpha}}}{i_1}{k_{\alpha}}}\\ \\
-C^{\bar{I}\bar{I} K}_{\bar{i}_1{\bar{i}_1}{k_1}}
\ea
\ea
\right)
\left(
\ba{c}
T^{\bar{I}i}_{\bar{\alpha}} \\ \\ T^{Kk}_{\bar{\alpha}}\\ \\T^{Ii}_{\bar{\alpha}}
\ea
\right),\label{case4}
\ee
where $X=C^{\overline{I}\overline{I} K}_{\bar{i}_1{\bar{i}_1}{k_1}}/C^{IIK}_{{i_1}{i_1}{k_1}}$.

(5) For $I,\overline{I},K$ contain Goldstones and $K$ is real, if
\be
W=m_I I\overline{I}+\frac{1}{2}m_K^2+\lambda I \overline{I}K,\nonumber
\ee
which gives
\be
0=\left(
\ba{ccc}
\ba{c}
-C^{I\bar{I}K}_{{i_1}{\bar{i}_1}{k_1}} \\ \\
C^{I\bar{I}K}_{{i_1}{\bar{i}_{\bar{\alpha}}}{k_{\alpha}}}\\ \\
0
\ea
\ba{c}
C^{I\bar{I}K}_{{i_{\alpha}}{\bar{i}_1}{k_{\bar{\alpha}}}}\\ \\
-C^{I\bar{I}K}_{{i_1}{\bar{i}_1}{k_1}} \\ \\
C^{I\bar{I}K}_{{i_1}{\bar{i}_{\alpha}}{k_{\bar{\alpha}}}}
\ea
\ba{c}
0\\ \\
C^{I\bar{I}K}_{i_{\bar{\alpha}} {\bar{i}_1}{k_{\alpha}}}\\ \\
-C^{I\bar{I}K}_{{i_1}{\bar{i}_1}{k_1}}
\ea
\ea
\right)
\left(
\ba{c}
T^{\bar{I}i}_{\bar{\alpha}} \\ \\ T^{Kk}_{\bar{\alpha}}\\ \\T^{Ii}_{\bar{\alpha}}
\ea
\right).\label{case5}
\ee
Furthermore, in case that
$I_i$ ($\overline{I}$) contains only $\alpha$ ($\bar{\alpha}$) but not $\bar{\alpha}$ ($\alpha$), we have
\be
0=\left(
\ba{cc}
\ba{c}
-C^{I\bar{I}K}_{{i_1}{\bar{i}_1}{k_1}} \\ \\
C^{I\bar{I}K}_{{i_1}{\bar{i}_{\bar{\alpha}}}{k_{\alpha}}}
\ea
\ba{c}
C^{I\bar{I}K}_{{i_{\alpha}}{\bar{i}_1}{k_{\bar{\alpha}}}}\\ \\
-C^{I\bar{I}K}_{{i_1}{\bar{i}_1}{k_1}}
\ea
\ea
\right)
\left(
\ba{c}
T^{\bar{I}i}_{\bar{\alpha}} \\ \\ T^{Kk}_{\bar{\alpha}}
\ea
\right).\label{case6}
\ee

We summarize in this Section that the Goldstino and thus the Goldstone components
are proportional to the VEVs,
and the remaining determinations  of these components need only the calculations of the CGCs.
This is the approach which we will use to determine the Goldstones in the following Sections,
and the identities of the CGC determinants will be used to check the consistencies of
these calculations.

\section{SSB and Goldstones in SO(10)}

We study the most general renormalizable couplings containing Higgs
$H(10)$, $D(120)$, $\overline{\Delta}(\overline{126})+ \Delta(126)$, $A(45)$,
$E(54)$ and $\Phi(210)$ in the SUSY $SO(10)$ models.
The most general renormalizable Higgs superpotential is \cite{fuku}
\bea
W &=& \frac{1}{2} m_{1} \Phi^2 + m_{2} \overline{\Delta} \Delta + \frac{1}{2} m_{3} H^2+ \frac{1}{2} m_{4} A^2 + \frac{1}{2} m_{5} E^2 + \frac{1}{2} m_{6} D^2
\nonumber\\
&+& \lambda_{1} \Phi^3 + \lambda_{2} \Phi \overline{\Delta} \Delta
+ \left(\lambda_3 \Delta + \lambda_4 \overline{\Delta} \right) H \Phi
+\lambda_{5} A^2 \Phi -i \lambda_{6} A \overline{\Delta} \Delta
+ \frac{\lambda_7}{120} \varepsilon A \Phi^2
\nonumber\\
&+& E \left( \lambda_{8} E^2 + \lambda_{9} A^2 + \lambda_{10} \Phi^2
+ \lambda_{11} \Delta^2 + \lambda_{12} \overline{\Delta}^2 + \lambda_{13} H^2
\right)
+D^2 \left( \lambda_{14} E + \lambda_{15} \Phi \right)
\nonumber\\
&+& D \left\{ \lambda_{16} H A + \lambda_{17} H \Phi + \left(
\lambda_{18} \Delta + \lambda_{19} \overline{\Delta} \right) A
+ \left( \lambda_{20} \Delta + \lambda_{21} \overline{\Delta} \right) \Phi
\right\},
\label{potential}
\eea

When we study the SSB of a SO(10) subgroup $G_1$ into $G_2$,
we firstly decompose the SO(10) representations into  $G_1$ representations,
and the couplings are now of the form
\be
\lambda^{IJK} IJK=\lambda^{IJK}\sum_{ijk}  C^{IJK}_{ijk} I_i J_j K_k,\label{coupg1}
\ee
where, for example, $I_i$ stands for the representations $i$ of $G_1$ from SO(10) superfield $I$.
Then, under $G_2$, $i,j,k$ are further decomposed, (\ref{coupg1}) is
\bea
\lambda^{IJK} I J K&=&\lambda^{IJK}\sum_{ijk}\sum_{abc}  C^{IJK}_{ijk}C^{ijk}_{abc} I_{i_a} J_{j_b} K_{kc}\nonumber\\
&\equiv& \lambda^{IJK}\sum_{ijk}\sum_{abc}  C^{I_i J_j K_k}_{a b c} I_{i_a} J_{j_b} K_{k_c},\label{coupg2}
\eea
where $I_{i_a}$ is a representations $a$ of $G_2$ coming from $I_i$, and
$C^{IJK}_{i_a j_b k_c}$ is the $SO(10)$ CGC.
Most of these CGCs have been given \cite{fuku,czy} so that no separate calculations of $C^{IJK}_{ijk}$ and $C^{ijk}_{abc}$ are needed.
A special kind of CGCs relevant for the SSB of the flipped $SU(5)$ symmetry will be given later.

To study SSB of $SO(10)$ into $G_{321}$,
it is necessary to decompose the Higgs representations of $SO(10)$ under the $G_{321}$ subgroup.
There are $45 - 12 = 33$ Goldstone modes.
We link the Goldstones of a SM representation with the SSB of
a specific subgroup of SO(10), as is summarized in Table \ref{tabsymm}.
In Table \ref{tabsymm}, the first three modes can be studied by the SO(10) CGCs using
$G_{422}$ as the maximal subgroup\cite{fuku},
the  $(3,2, -\frac{5}{6})+{\rm c.c.}$ modes need the CGCs using $SU(5)$\cite{czy},
and $(3,2,\frac{1}{6})+{\rm c.c.}$ using $G_{51}$ CGCs which
will be given below.

\section{Goldstone modes for SSB of $G_{422}$}

Under $G_{422}$, the following representations
\bea
A_1\equiv \widehat{A}_{(1,1,3)}, ~~
A_2\equiv \widehat{A}_{(15,1,1)}, ~~
E \equiv \widehat{E}_{(1,1,1)},\nonumber\\
v_R\equiv \widehat{\Delta}_{(\overline{10},1,3)},~~
\overline{v_R}\equiv \widehat{\overline{\Delta}}_{(10,1,3)},~~~~~~~~~~~ \nonumber\\
\Phi_1\equiv \widehat{\Phi}_{(1,1,1)},~~
\Phi_2\equiv \widehat{\Phi}_{(15,1,1)},~~
\Phi_3\equiv \widehat{\Phi}_{(15,1,3)},\label{422VEV}
\eea
contain the SM singlets $(1,1,0)$ whose VEVs will be denoted in the same symbols.
Obviously, $E$ and $\Phi_1$ contain no Goldstones relevant for the $G_{422}$ breaking,
$A_1$ contains Goldstone components relevant for the $SU(2)_R$ breaking,
$A_2,\Phi_2$ for the $SU(4)_C$ breaking, and $v_R,\overline{v_R},\Phi_3$ for both.
The needed  CGCs are taken from \cite{fuku} and are summarized in Table \ref{tab110},\ref{tab111},\ref{tab31}.

\subsection{Goldstone (1,1,0)  under $G_{321}$}

This Goldstone is the result of the SSB of $U(1)_{I_{3R}}\otimes U(1)_{B-L}$ into $U(1)_Y$
and has been studied in \cite{rzx}.
In the renormalizable models, only $\Delta(\textbf{126}) +\overline{\Delta}(\overline{\textbf{126}})$
have the SM singlets with nonzero $U(1)_{I_{3R}},U(1)_{B-L}$ charges. Furthermore, if we include also fields
in the spinor representations $\Psi(\textbf{16})+\overline{\Psi}(\overline{\textbf{16}})$ of SO(10), which are
usually used in building  non-renormalizable models and have halves
of the charges of $\Delta(\textbf{126}) +\overline{\Delta}(\overline{\textbf{126}})$,
the Goldstino and thus the Goldstone mode is found to be
\be
\frac{\langle\Delta\rangle}{N} \widehat{\Delta}^{(1,1,0)}_{(\overline{10},1,3)}
-\frac{\langle\overline{\Delta}\rangle}{N} \widehat{\overline{\Delta}}^{(1,1,0)}_{(10,1,3)}
+\frac{1}{2} \frac{\langle\Psi\rangle}{N} \widehat{\Psi}^{(1,1,0)}_{(\overline{4},1,2)}
-\frac{1}{2} \frac{\langle\overline{\Psi}\rangle}{N}\widehat{\overline{\Psi}}^{(1,1,0)}_{(4,1,2)},\label{u1gold}
\ee
where $N$ is a simple normalization factor.
Equation (\ref{u1gold}) tell us two important conclusions in the U(1) symmetry breaking.
First, the component of a field in the Goldstone mode is proportional to its charge
under the breaking $U(1)$ and to its VEV.
Second, there is no dependence on the superpotential parameters of the model,
besides through the VEV determinations indirectly.

In the  model (\ref{potential}), we have
\begin{equation}
	\overrightarrow{G}_{(1,1,0)} = v_R \widehat{\Delta}^{(1,1,0)}_{(\overline{10},1,3)}
	- \overline{v_R} \widehat{\overline{\Delta}}^{(1,1,0)}_{(10,1,3)}.
\end{equation}
up to an obvious normalization.

\subsection{[(1,1,1) + c.c.]}

They are relevant for the SSB of $SU(2)_R$ into $U(1)_{I_{3R}}$.
The fields contain $\bar{\alpha}=(1,1,1)$ are
$$\widehat{A}^{(1,1,1)}_{(1,1,3)},
\widehat{D}^{(1,1,1)}_{(\overline{10},1,1)},
\widehat{\Delta}^{(1,1,1)}_{(\overline{10},1,3)},
\widehat{\Phi}^{(1,1,1)}_{(15,1,3)}.$$
Note that $D$ has no VEV, the Goldstone corresponding to $\overline{\alpha}=(1,1,1)$ of $G_{321}$ is
written as
\begin{equation}
\overrightarrow{G}_{(1,1,1)}
= T^{A_1}_{\overline{\alpha}} A_1 \widehat{A}^{(1,1,1)}_{(1,1,3)}
+ T^{v_R}_{\overline{\alpha}} v_R \widehat{\Delta}^{(1,1,1)}_{(\overline{10},1,3)}
+ T^{\Phi_3}_{\overline{\alpha}} \Phi_3 \widehat{\Phi}^{(1,1,1)}_{(15,1,3)}.\nonumber
\end{equation}
Here $A_1\dots$ are the VEVs, $\widehat{A},\dots$ are the fields and $T$s are what will be solved.
We will first classify the couplings and discuss their results, both on solve the Goldstones
and on the identities among the CGCs.

1) The representations $E,A_2,\Phi_1,\Phi_2$ have no $(1,1,\pm 1)$ under $G_{321}$,
so that couplings of them with two same representations, such as $A_1^2 E$ or $\Phi_3^2 \Phi_1$,
do not give relations among the $T$s. These couplings lead to relations of CGCs such as
\bea
	0&=&	 C^{A_1 A_1 E}_{1~1~1} - C^{A_1 A_1 E}_{\bar{\alpha}~\alpha ~1}
= \frac{\sqrt{3}}{2\sqrt{5}}
- \frac{\sqrt{3}}{2\sqrt{5}},\nonumber\\
0&=& C^{\Phi_3 \Phi_3 \Phi_1}_{1~1~1}
- C^{\Phi_3 \Phi_3 \Phi_1}_{\bar{\alpha}~\alpha ~1}
=\frac{1}{6\sqrt{6}}-\frac{1}{6\sqrt{6}},\nonumber
\eea
which are trivial.

2) Any coupling of $E,A_2,\Phi_1,\Phi_2$ with two different representations will set up
a relation of the two $T$s. Following (\ref{case1}),
the coupling $A_1\Phi_3 A_2$ gives
\begin{equation}
0 = \left( \begin{array}{cc}
-C^{A_1 \Phi_3 A_2}_{1~1~1}  & C^{A_1 \Phi_3 A_2}_{\alpha~\bar{\alpha}~1}   \\ \\
C^{A_1 \Phi_3 A_2}_{\bar{\alpha}~\alpha~1} & - C^{A_1 \Phi_3 A_2}_{1~1~1}
\end{array}
\right)
\left(
\begin{array}{c}
T^{A_1}_{\bar{\alpha}} \\ \\ T^{\Phi_3}_{\bar{\alpha}}
\end{array}
\right)
= \left( \begin{array}{cc}
- \dfrac{1}{\sqrt{6}}  & - \dfrac{1}{\sqrt{6}}    \\ \\
- \dfrac{1}{\sqrt{6}}
& - \dfrac{1}{\sqrt{6}}
\end{array}
\right)
\left(
\begin{array}{c}
T^{\Phi_2}_{\bar{\alpha}} \\ \\ T^{\Phi_3}_{\bar{\alpha}}
\end{array}
\right), \nonumber
\end{equation}
where the determinant is zero, so
\be
\frac{T^{A_1}_{\bar{\alpha}}}{T^{\Phi_3}_{\bar{\alpha}}} = \frac{C^{A_1 \Phi_3 A_2}_{\bar{\alpha}~\alpha~1}}{C^{A_1 \Phi_3 A_2}_{1~1~1}} = (- \frac{1}{\sqrt{6}}) / (\frac{1}{\sqrt{6}}) = - 1.\nonumber
\ee
The coupling $A_1 \Phi_3 \Phi_2$ also gives
\begin{equation}
	\frac{T^{A_1}_{\bar{\alpha}}}{T^{\Phi_3}_{\bar{\alpha}}} = \frac{C^{A_1 \Phi_3 \Phi_2}_{\bar{\alpha}~\alpha~1}}{C^{A_1 \Phi_3 \Phi_2}_{1~1~1}}  = (- \frac{1}{\sqrt{6}}) / (\frac{1}{\sqrt{6}}) = -1.
\nonumber
\end{equation}

3) $\Phi_3 v_R \overline{v_R}$ and $A_1 v_R \overline{v_R}$ are couplings with all three
representations contain the Goldstone modes. However, $v_R$ contains only $\overline{\alpha}$ while $\overline{v_R}$
contains only $\alpha$. Following (\ref{case6}),
the coupling $\Phi_3 v_R \overline{v_R}$ gives
\begin{equation}
	\frac{T^{\Phi_3}_{\bar{\alpha}}}{T^{v_R}_{\bar{\alpha}}} = \frac{C^{v_R \Phi_3  \overline{v_R}}_{\bar{\alpha}~\alpha~1}}{C^{v_R \Phi_3 \overline{v_R}}_{1~1~1}}
	= (- \frac{1}{10}) / (\frac{1}{10}) = - 1,\nonumber
\end{equation}
and $A_1 v_R \overline{v_R}$ gives
\begin{equation}
	\frac{T^{A_1}_{\bar{\alpha}}}{T^{v_R}_{\bar{\alpha}}} = \frac{C^{v_R A_1  \overline{v_R}}_{\bar{\alpha}~\alpha~1}}{C^{v_R A_1  \overline{v_R}}_{1~1~1} }
	= (- \frac{1}{5}) / (- \frac{1}{5}) = 1.\nonumber
\end{equation}

Altogether, we have consistently
$T^{A_1}_{\bar{\alpha}}: T^{v_R}_{\bar{\alpha}}:T^{\Phi_3}_{\bar{\alpha}} = 1:1: (-1)$,
so the Goldstone mode is
\begin{equation}
\overrightarrow{G}_{(1,1,1)}
=  A_1 \widehat{A}^{(1,1,1)}_{(1,1,3)}
+  v_R \widehat{\Delta}^{(1,1,1)}_{(\overline{10},1,3)}
- \Phi_3 \widehat{\Phi}^{(1,1,1)}_{(15,1,3)}
\end{equation}
with an obvious normalization.

\subsection{[(3,1,$\frac{2}{3}$) + c.c.]}

They are the Goldstones for $SU(4)_C$ SSB into $SU(3)_C\otimes U(1)_{B-L}$.
The fields corresponding to $\bar{\alpha}=(3,1,\frac{2}{3})$ are
$$\widehat{A}^{(3,1,\frac{2}{3})}_{(15,1,1)},
\widehat{D}^{(3,1,\frac{2}{3})}_{(6,1,3)},
\widehat{\overline{\Delta}}^{(3,1,\frac{2}{3})}_{(10,1,3)},
\widehat{\Phi}^{(3,1,\frac{2}{3})}_{(15,1,1)},
\widehat{\Phi}^{(3,1,\frac{2}{3})}_{(15,1,3)}.$$
Again, $D$ does not have a VEV, the Goldstone mode is
\begin{equation}
	\overrightarrow{G}_{(3,1,\frac{2}{3})}
	= T^{A_2}_{\bar{\alpha}} A_2 \widehat{A}^{(3,1,\frac{2}{3})}_{(15,1,1)}
	+ T^{\overline{v_R}}_{\bar{\alpha}} \overline{v_R} \widehat{\overline{\Delta}}^{(3,1,\frac{2}{3})}_{(10,1,3)}
	+ T^{\Phi_2}_{\bar{\alpha}} \Phi_2 \widehat{\Phi}^{(3,1,\frac{2}{3})}_{(15,1,1)}
	+ T^{\Phi_3}_{\bar{\alpha}} \Phi_3 \widehat{\Phi}^{(3,1,\frac{2}{3})}_{(15,1,3)}.\nonumber
\end{equation}

1) According to (\ref{case3}), $\Phi_2^3$ leads to
\begin{eqnarray}
	0 = C^{\Phi_2 \Phi_2 \Phi_2}_{1~1~1} - 2 C^{\Phi_2 \Phi_2 \Phi_2}_{\bar{\alpha}~\alpha~1}= \frac{1}{9\sqrt{2}} - 2\ast \frac{1}{18\sqrt{2}}.\nonumber
\end{eqnarray}

2) $A_1,E,\Phi_1$ does not contain $\alpha$ or $\overline{\alpha}$ and are SU(4) singlets,
their couplings with the same
fields, $A_2^2 E$, $\Phi_3^2 E$, $\Phi_3^2 \Phi_1$ lead to only trivial results.
In the couplings of one of them with two different fields, according to (\ref{case1}),
they give relations between two Goldstone components.
$A_2 \Phi_3 A_1$ gives
\begin{equation}
	\frac{T^{A_2}_{\bar{\alpha}}}{T^{\Phi_3}_{\bar{\alpha}}} = \frac{C^{A_2 \Phi_3 A_1}_{\bar{\alpha}~\alpha~1}}{C^{A_2 \Phi_3 A_1}_{1~1~1}} = (- \frac{1}{\sqrt{6}}) / (\frac{1}{\sqrt{6}}) = - 1,\nonumber
\end{equation}
$A_2 \Phi_2 \Phi_1$ gives
\begin{equation}
	\frac{T^{A_2}_{\bar{\alpha}}}{T^{\Phi_2}_{\bar{\alpha}}} = \frac{C^{A_2 \Phi_2 \Phi_1}_{\bar{\alpha}~\alpha~1}}{C^{A_2 \Phi_2 \Phi_1}_{1~1~1}} = (- \frac{\sqrt{2}}{5}) / (\frac{\sqrt{2}}{5}) = - 1,
\nonumber
\end{equation}
and $\Phi_2 \Phi_3 A_1$ gives
\begin{equation}
	\frac{T^{A_2}_{\bar{\alpha}}}{T^{\Phi_3}_{\bar{\alpha}}}
= \frac{C^{A_2 \Phi_3 A_1}_{\bar{\alpha}~\alpha~1}}{C^{A_2 \Phi_3 A_1}_{1~1~1}} = (- \frac{1}{\sqrt{6}}) / (\frac{1}{\sqrt{6}}) = - 1.\nonumber
\end{equation}

3) $v_R$ contains only $\overline{\alpha}$ while $\overline{v_R}$ contains only ${\alpha}$.
According to (\ref{case6}),
$ v_R \overline{v_R}\Phi_2$ gives
\begin{equation}
	\frac{T^{\Phi_2}_{\bar{\alpha}}}{T^{\overline{v_R}}_{\bar{\alpha}}} = \frac{C^{ \overline{v_R} \Phi_2 v_R}_{\bar{\alpha}~\alpha~1}}{C^{\overline{v_R} \Phi_2  v_R}_{1~1~1}}
	= (-\frac{1}{10\sqrt{3}}) / (\frac{1}{10\sqrt{2}}) = - \sqrt{\frac{2}{3}},\nonumber
\end{equation}
$v_R \overline{v_R}\Phi_3 $ gives
\begin{equation}
\frac{T^{\Phi_3}_{\bar{\alpha}}}{T^{\overline{v_R}}_{\bar{\alpha}}} = \frac{C^{ \overline{v_R} \Phi_3 v_R}_{\bar{\alpha}~\alpha~1}}{C^{\overline{v_R} \Phi_3  v_R}_{1~1~1}}
= (-\frac{1}{5\sqrt{6}}) / (\frac{1}{10}) = - \sqrt{\frac{2}{3}},\nonumber
\end{equation}
and $v_R \overline{v_R}A_2 $ gives
\begin{equation}
	\frac{T^{A_2}_{\bar{\alpha}}}{T^{\overline{v_R}}_{\bar{\alpha}}} = \frac{C^{ \overline{v_R} A_2 v_R}_{\bar{\alpha}~\alpha~1}}{C^{\overline{v_R} A_2  v_R}_{1~1~1}}
	= (-\frac{1}{5}) / (- \frac{\sqrt{3}}{5\sqrt{2}}) = \sqrt{\frac{2}{3}}.\nonumber
\end{equation}

4) According to (\ref{case2}), $\Phi_2\Phi_3^2 $ gives
\begin{equation}
0 = \left( \begin{array}{cc}
-C^{\Phi_2 \Phi_3 \Phi_3}_{1~1~1}  & 2 C^{\Phi_2 \Phi_3 \Phi_3}_{\bar{\alpha}~\alpha~1}   \\ \\
C^{\Phi_3 \Phi_2 \Phi_3}_{\bar{\alpha}~\alpha~1}
& C^{\Phi_3 \Phi_3 \Phi_2}_{\bar{\alpha}~\alpha~1}
- C^{\Phi_3 \Phi_3 \Phi_2}_{1~1~1}
\end{array}
\right)
\left(
\begin{array}{c}
T^{\Phi_2}_{\bar{\alpha}} \\ \\ T^{\Phi_3}_{\bar{\alpha}}
\end{array}
\right)
= \left( \begin{array}{cc}
- \dfrac{1}{9\sqrt{2}}  & 2\ast\dfrac{1}{18\sqrt{2}}   \\ \\
\dfrac{1}{18\sqrt{2}}
& \dfrac{1}{9\sqrt{2}}
- \dfrac{1}{18\sqrt{2}}
\end{array}
\right)
\left(
\begin{array}{c}
T^{\Phi_2}_{\bar{\alpha}} \\ \\ T^{\Phi_3}_{\bar{\alpha}}
\end{array}
\right),\nonumber
\end{equation}
so
\begin{eqnarray}
	\frac{T_{\Phi_3}}{T_{\Phi_2}} &=& \frac{C^{\Phi_3 \Phi_2 \Phi_3}_{\bar{\alpha}~\alpha~1}}{C^{\Phi_3 \Phi_3 \Phi_2}_{1~1~1} - C^{\Phi_3 \Phi_3 \Phi_2}_{\bar{\alpha}~\alpha~1}} = (\frac{1}{18\sqrt{2}}) / (\frac{1}{9\sqrt{2}} - \frac{1}{18\sqrt{2}}) = 1. \nonumber\\
	or: \frac{T_{\Phi_3}}{T_{\Phi_2}} &=& \frac{C^{\Phi_2 \Phi_3 \Phi_3}_{1~1~1}}{2 C^{\Phi_2 \Phi_3 \Phi_3}_{\bar{\alpha}~\alpha~1}} = (\frac{1}{9\sqrt{2}}) / (2*\frac{1}{18\sqrt{2}}) = 1.\nonumber
\end{eqnarray}
Similarly,
$\Phi_2 A_2^2 $ gives
\begin{eqnarray}
	\frac{T^{A_2}_{\bar{\alpha}}}{T^{\Phi_2}_{\bar{\alpha}}} &=& \frac{C^{A_2 \Phi_2 A_2}_{\bar{\alpha}~\alpha~1}}{C^{A_2 A_2 \Phi_2}_{1~1~1} - C^{A_2 A_2 \Phi_2}_{\bar{\alpha}~\alpha~1}} = (- \frac{1}{3\sqrt{2}}) /(\frac{\sqrt{2}}{3} - \frac{1}{3\sqrt{2}}) = - 1. \nonumber\\
	or:
	\frac{T^{A_2}_{\bar{\alpha}}}{T^{\Phi_2}_{\bar{\alpha}}} &=& \frac{C^{\Phi_2 A_2 A_2}_{1~1~1}}{2 C^{\Phi_2 A_2 A_2}_{\bar{\alpha}~\alpha~1}} = (\frac{\sqrt{2}}{3}) / (2*\frac{-1}{3\sqrt{2}}) = -1.\nonumber
\end{eqnarray}
$A_2 \Phi_3^2$ gives
\begin{eqnarray}
	\frac{T^{A_2}_{\bar{\alpha}}}{T^{\Phi_3}_{\bar{\alpha}}} &=& \frac{2 C^{A_2 \Phi_3 \Phi_3}_{\bar{\alpha}~\alpha~1}}{C^{A_2 \Phi_3 \Phi_3}_{1~1~1}} = (- \frac{2\ast\sqrt{2}}{5\sqrt{3}}) / (\frac{2\sqrt{2}}{5\sqrt{3}}) = -1, \nonumber\\
	or:
	\frac{T^{A_2}_{\bar{\alpha}}}{T^{\Phi_3}_{\bar{\alpha}}} &=& \frac{C^{\Phi_3 \Phi_3 A_2}_{1~1~1} - C^{\Phi_3 \Phi_3 A_2}_{\bar{\alpha}~\alpha~1}}{C^{\Phi_3 A_2 \Phi_3}_{\bar{\alpha}~\alpha~1}}
	= (\frac{2\sqrt{2}}{5\sqrt{3}} - \frac{\sqrt{2}}{5\sqrt{3}}) / (-\frac{\sqrt{2}}{5\sqrt{3}})
	= - 1.\nonumber
\end{eqnarray}

All together we have $T_A = \sqrt{\tfrac{2}{3}} T^{\overline{v_R}}_{\bar{\alpha}} = - T^{\Phi_2}_{\bar{\alpha}} = - T^{\Phi_3}_{\bar{\alpha}}$,
then the Goldstone is
\begin{equation}
\overrightarrow{G}_{(3,1,\frac{2}{3})}
=  A_2 \widehat{A}^{(3,1,\frac{2}{3})}_{(15,1,1)}
+ \sqrt{\frac{3}{2}} \overline{v_R} \widehat{\overline{\Delta}}^{(3,1,\frac{2}{3})}_{(10,1,3)}
- \Phi_2 \widehat{\Phi}^{(3,1,\frac{2}{3})}_{(15,1,1)}
- \Phi_3 \widehat{\Phi}^{(3,1,\frac{2}{3})}_{(15,1,3)}.
\end{equation}

\section{Goldstone modes for SSB of $SU(5)$: [(3,2,$-\frac{5}{6}$) + c.c.]}

In study SSB of $SU(5)$, the fields need to be decomposed into $SU(5)$ representations
and use the CGCs calculated in \cite{czy}. Under SU(5), the following representations
\begin{eqnarray}
{a_1} &\equiv& \widehat{A}_{(1)},~~~
{a_2} \equiv \widehat{A}_{(24)},~~~
{S} \equiv \widehat{E}_{(24)}, \nonumber\\
{V_R} &\equiv& \widehat{\Delta}_{(1)},~~~
{\overline{V_R}} \equiv \widehat{\overline{\Delta}}_{(1)},\nonumber\\
{\phi}_1 &\equiv& \widehat{\Phi}_{(1)},  ~~~
{\phi}_2 \equiv \widehat{\Phi}_{(24)}, ~~~
{\phi}_3 \equiv \widehat{\Phi}_{(75)}. \label{su5VEV}
\end{eqnarray}
contain the SM singlets whose VEVs will be denoted by the same symbols.
Here the subscripts are the representations under $SU(5)$.

Since the $SU(5)$ singlets do not contribute to the Goldstone modes,
the fields contains $\bar{\alpha}=(3,2,-\frac{5}{6})$ are
$$\widehat{A}^{(3,2,-\frac{5}{6})}_{(24)},
\widehat{E}^{(3,2,-\frac{5}{6})}_{(24)},
\widehat{\Phi}^{(3,2,-\frac{5}{6})}_{(24)},
\widehat{\Phi}^{(3,2,-\frac{5}{6})}_{(75)}.$$
The Goldstone corresponding  to $(3,2,-\frac{5}{6})$ is denoted as
\begin{equation}
	\overrightarrow{G}_{(3,2,-\frac{5}{6})}
	= T^{a_2}_{\bar{\alpha}} a_2 \widehat{A}^{(3,2,-\frac{5}{6})}_{(24)}
	+ T^{S}_{\bar{\alpha}} S \widehat{E}^{(3,2,-\frac{5}{6})}_{(24)}
	+ T^{\phi_2}_{\bar{\alpha}} \phi_2 \widehat{\Phi}^{(3,2,-\frac{5}{6})}_{(24)}
	+ T^{\phi_3}_{\bar{\alpha}} \phi_3 \widehat{\Phi}^{(3,2,-\frac{5}{6})}_{(75)}.\nonumber
\end{equation}
The needed CGCs are taken from \cite{czy} and summarized in Table \ref{tabsu5a},\ref{tabsu5b}.

1) According to (\ref{case3}), the coupling $\phi_2^3$ gives
\be
0 = C^{\phi_2 \phi_2 \phi_2}_{1~1~1} - 2 C^{\phi_2 \phi_2 \phi_2}_{\bar{\alpha}~\alpha~1}
=(- \frac{7}{9\sqrt{15}}) - 2\ast (- \frac{7}{18\sqrt{15}}),\nonumber
\ee
$\phi_3^3$ gives
\begin{eqnarray}
0 =C^{\phi_3 \phi_3 \phi_3}_{1~1~1} - 2 C^{\phi_3 \phi_3 \phi_3}_{\bar{\alpha}~\alpha~1}
= \frac{8}{9\sqrt{3}} - 2\ast\frac{4}{9\sqrt{3}},\nonumber
\end{eqnarray}
and $S^3$ gives
\be
0 =C^{SSS}_{111} - 2 C^{SSS}_{\bar{\alpha} \alpha 1} =\sqrt{\frac{3}{5}} - 2 *\frac{\sqrt{3}}{2\sqrt{5}}.\nonumber
\ee

2) $\phi_1, a_1$ contain neither Goldstones, the couplings
$\phi_2^2 \phi_1$, $\phi_3^2 \phi_1$, $a_2^2 \phi_1$, $\phi_2^2 a_1$ and $\phi_3^2 a_1$
lead to only trivial results.
The coupling $a_2 \phi_2 a_1$, according to (\ref{case1}), gives
\be
\frac{T^{a_2}_{\bar{\alpha}}}{T^{\phi_2}_{\bar{\alpha}}} = \frac{C^{a_2 \phi_2 a_1}_{\bar{\alpha}~\alpha~1}}{C^{a_2 \phi_2 a_1}_{1~1~1}} = (-\sqrt{\frac{2}{5}}) / (\sqrt{\frac{2}{5}}) = -1,\nonumber
\ee
$a_2 \phi_2 \phi_1$ gives
\begin{equation}
\frac{T^{a_2}_{\bar{\alpha}}}{T^{\phi_2}_{\bar{\alpha}}} = \frac{C^{a_2 \phi_2 \phi_1}_{\bar{\alpha}~\alpha~1}}{C^{a_2 \phi_2 \phi_1}_{1~1~1}} = (\frac{2\sqrt{3}}{5\sqrt{5}}) / (-\frac{2\sqrt{3}}{5\sqrt{5}})
= -1,\nonumber
\end{equation}
$S a_2 a_1$ gives
\begin{equation}
\frac{T^{S}_{\bar{\alpha}}}{T^{a_2}_{\bar{\alpha}}} = \frac{C^{S a_2 a_1}_{\bar{\alpha}~\alpha~1}}{C^{S a_2 a_1}_{1~1~1}}
= (-\sqrt{\frac{2}{5}}) / (\sqrt{\frac{2}{5}}) = -1,\nonumber
\end{equation}
and $S \phi_2 \phi_1$ gives
\begin{equation}
\frac{T^{S}_{\bar{\alpha}}}{T^{\phi_2}_{\bar{\alpha}}} = \frac{C^{S \phi_2 \phi_1}_{\bar{\alpha}~\alpha~1}}
{C^{S \phi_2 \phi_1}_{1~1~1}} = (\frac{\sqrt{3}}{2\sqrt{5}} ) / (\frac{\sqrt{3}}{2\sqrt{5}} ) = 1.\nonumber
\end{equation}

3) According to (\ref{case2}), $\phi_2 \phi_3^2$ gives
\begin{eqnarray}
	\frac{T^{\phi_2}_{\bar{\alpha}}}{T^{\phi_3}_{\bar{\alpha}}}
	&=& \frac{2 C^{\phi_2 \phi_3 \phi_3}_{\bar{\alpha}~\alpha~1}}{C^{\phi_2 \phi_3 \phi_3}_{1~1~1}}
	= 2* (- \frac{\sqrt{2}}{9\sqrt{3}}) / (-\frac{8}{9\sqrt{15}})
	= \sqrt{\frac{5}{8}}, \nonumber\\
	or:
	\frac{T^{\phi_2}_{\bar{\alpha}}}{T^{\phi_3}_{\bar{\alpha}}}
	&=& \frac{C^{\phi_3 \phi_3 \phi_2}_{1~1~1} - C^{\phi_3 \phi_3 \phi_2}_{\bar{\alpha}~\alpha~1}}{C^{\phi_3 \phi_2 \phi_3}_{\bar{\alpha}~\alpha~1}}
	= (\frac{-8}{9\sqrt{15}} - \frac{-11}{18\sqrt{15}}) / (- \frac{\sqrt{2}}{9\sqrt{3}})
	= \sqrt{\frac{5}{8}},\nonumber
\end{eqnarray}
$\phi_2^2 \phi_3$ gives
\begin{eqnarray}
	\frac{T^{\phi_2}_{\bar{\alpha}}}{T^{\phi_3}_{\bar{\alpha}}}
	&=& \frac{C^{\phi_2 \phi_3 \phi_2}_{\bar{\alpha}~\alpha~1}}{C^{\phi_2 \phi_2 \phi_3}_{1~1~1} - C^{\phi_2 \phi_2 \phi_3}_{\bar{\alpha}~\alpha~1}}
	= (\frac{\sqrt{5}}{9\sqrt{6}}) / (\frac{5}{18\sqrt{3}} - \frac{1}{18\sqrt{3}})
	= \sqrt{\frac{5}{8}}, \nonumber\\
	or:
	\frac{T^{\phi_2}_{\bar{\alpha}}}{T^{\phi_3}_{\bar{\alpha}}}
	&=& \frac{C^{\phi_3 \phi_2 \phi_2}_{1~1~1} }{2 C^{\phi_3 \phi_2 \phi_2}_{\bar{\alpha}~\alpha~1}}
	= (\frac{5}{18\sqrt{3}}) / (2*\frac{\sqrt{5}}{9\sqrt{6}})
	= \sqrt{\frac{5}{8}},\nonumber
\end{eqnarray}
$a_2^2 \phi_2$ gives
\begin{eqnarray}
\frac{T^{a_2}_{\bar{\alpha}}}{T^{\phi_2}_{\bar{\alpha}}}
&=& \frac{C^{a_2 \phi_2 a_2}_{\bar{\alpha}~\alpha~1}}{C^{a_2 a_2 \phi_2}_{1~1~1} - C^{a_2 a_2 \phi_2}_{\bar{\alpha}~\alpha~1}}
= (\frac{1}{3\sqrt{15}}) / (\frac{-2}{3\sqrt{15}} - \frac{-1}{3\sqrt{15}}) = -1, \nonumber\\
or:
\frac{T^{a_2}_{\bar{\alpha}}}{T^{\phi_2}_{\bar{\alpha}}}
&=& \frac{C^{\phi_2 a_2 a_2}_{1~1~1}}{2 C^{\phi_2 a_2 a_2 }_{\bar{\alpha}~\alpha~1}}
= (-\frac{2}{3\sqrt{15}}) / 2*(\frac{1}{3\sqrt{15}}) = -1,\nonumber
\eea
$a_2^2 \phi_3$ gives
\begin{eqnarray}
\frac{T^{a_2}_{\bar{\alpha}}}{T^{\phi_3}_{\bar{\alpha}}}
&=& \frac{C^{a_2 \phi_3 a_2}_{\bar{\alpha}~\alpha~1}}{C^{a_2 a_2 \phi_3}_{1~1~1} - C^{a_2 a_2 \phi_3}_{\bar{\alpha}~\alpha~1}}
= (- \frac{\sqrt{10}}{3\sqrt{3}}) / (\frac{5}{3\sqrt{3}} - \frac{1}{3\sqrt{3}}) = -\sqrt{\frac{5}{8}}, \nonumber\\
or:
\frac{T^{a_2}_{\bar{\alpha}}}{T^{\phi_3}_{\bar{\alpha}}}
&=& \frac{C^{\phi_3 a_2 a_2}_{1~1~1}}{2 C^{\phi_3 a_2 a_2 }_{\bar{\alpha}~\alpha~1}}
= (\frac{5}{3\sqrt{3}}) / (2*(-\frac{\sqrt{10}}{3\sqrt{3}})) = -\sqrt{\frac{5}{8}},\nonumber
\end{eqnarray}
$\phi_2^2 a_2$ gives
\begin{eqnarray}
\frac{T^{a_2}_{\bar{\alpha}}}{T^{\phi_2}_{\bar{\alpha}}} &=& \frac{2 C^{a_2 \phi_2 \phi_2}_{\bar{\alpha}~\alpha~1}}{C^{a_2 \phi_2 \phi_2}_{1~1~1}} = 2*(-\frac{4}{15\sqrt{15}}) /(\frac{8}{15\sqrt{15}}) = -1, \notag\\
or:
\frac{T^{a_2}_{\bar{\alpha}}}{T^{\phi_2}_{\bar{\alpha}}} &=& \frac{C^{\phi_2 \phi_2 a_2}_{1~1~1}
- C^{\phi_2 \phi_2 a_2}_{\bar{\alpha}~\alpha~1} }
{C^{\phi_2 a_2 \phi_2}_{\bar{\alpha}~\alpha~1}}
= (\frac{8}{15\sqrt{15}}- \frac{4}{15\sqrt{15}}) / (- \frac{4}{15\sqrt{15}}) = -1,\nonumber
\end{eqnarray}
$\phi_3^2 a_2$ gives
\begin{eqnarray}
\frac{T^{a_2}_{\bar{\alpha}}}{T^{\phi_3}_{\bar{\alpha}}} &=& \frac{2 C^{a_2 \phi_3 \phi_3}_{\bar{\alpha}~\alpha~1}}{C^{a_2 \phi_3 \phi_3}_{1~1~1}} = 2*(\frac{4\sqrt{2}}{15\sqrt{3}}) /(- \frac{32}{15\sqrt{15}}) = -\sqrt{\frac{5}{8}}, \notag\\
or:
\frac{T^{a_2}_{\bar{\alpha}}}{T^{\phi_3}_{\bar{\alpha}}} &=& \frac{C^{\phi_3 \phi_3 a_2}_{1~1~1}
	- C^{\phi_3 \phi_3 a_2}_{\bar{\alpha}~\alpha~1} }
{C^{\phi_3 a_2 \phi_3}_{\bar{\alpha}~\alpha~1}}
= (\frac{-32}{15\sqrt{15}} - \frac{-22}{15\sqrt{15}}) / (\frac{4\sqrt{2}}{15\sqrt{3}}) = -\sqrt{\frac{5}{8}},\nonumber
\end{eqnarray}
$S a_2^2$ gives
\begin{eqnarray}
\frac{T^{S}_{\bar{\alpha}}}{T^{a_2}_{\bar{\alpha}}} &=&
\frac{2 C^{S a_2 a_2}_{\bar{\alpha}~\alpha~1}}{C^{S a_2 a_2}_{1~1~1}}
= 2*(- \frac{1}{2\sqrt{15}} ) / (\frac{1}{\sqrt{15}} ) = -1, \nonumber\\
or:
\frac{T^{S}_{\bar{\alpha}}}{T^{a_2}_{\bar{\alpha}}} &=&
\frac{C^{a_2 a_2 S}_{1~1~1} - C^{a_2 a_2 S}_{\bar{\alpha}~\alpha~1}}
{C^{a_2 S a_2}_{\bar{\alpha}~\alpha~1}}
= \left(\frac{1}{\sqrt{15}} - \frac{1}{2\sqrt{15}}\right)/\left(-\frac{1}{2\sqrt{15}}\right) = -1,\nonumber
\end{eqnarray}
$S \phi_2^2$ gives
\begin{eqnarray}
\frac{T^{S}_{\bar{\alpha}}}{T^{\phi_2}_{\bar{\alpha}}} &=&
\frac{2 C^{S \phi_2 \phi_2}_{\bar{\alpha}~\alpha~1}}
{C^{S \phi_2 \phi_2}_{1~1~1}} = 2*(\frac{1}{12\sqrt{15}}) / (\frac{1}{6\sqrt{15}}) = 1, \nonumber\\
or:
\frac{T^{S}_{\bar{\alpha}}}{T^{\phi_2}_{\bar{\alpha}}} &=&
\frac{C^{\phi_2 \phi_2 S}_{1~1~1} - C^{\phi_2 \phi_2 S}_{\bar{\alpha}~\alpha~1}}
{C^{\phi_2 S \phi_2}_{\bar{\alpha}~\alpha~1}}
= \left(\frac{1}{6\sqrt{15}} - \frac{1}{12\sqrt{15}}\right)/\left(\frac{1}{12\sqrt{15}}\right) = 1,\nonumber
\end{eqnarray}
and  $S \phi_3^2$ gives
\begin{eqnarray}
\frac{T^{S}_{\bar{\alpha}}}{T^{\phi_3}_{\bar{\alpha}}} &=&
\frac{2 C^{S \phi_3 \phi_3}_{\bar{\alpha}~\alpha~1}}
{C^{S \phi_3 \phi_3}_{1~1~1}} = 2*(\frac{1}{3\sqrt{6}}) / (\frac{4}{3\sqrt{15}}) = \sqrt{\frac{5}{8}}, \nonumber\\
or:
\frac{T^{S}_{\bar{\alpha}}}{T^{\phi_3}_{\bar{\alpha}}} &=&
\frac{C^{\phi_3 \phi_3 S}_{1~1~1} - C^{\phi_3 \phi_3 S}_{\bar{\alpha}~\alpha~1}}
{C^{\phi_3 S \phi_3}_{\bar{\alpha}~\alpha~1}}
= \left(\frac{4}{3\sqrt{15}} - \frac{11}{12\sqrt{15}}\right)/\left(\frac{1}{3\sqrt{6}}\right)
= \sqrt{\frac{5}{8}}.\nonumber
\end{eqnarray}

4) According to (\ref{19}), $a_2 \phi_2 \phi_3$ gives
\be
0 = \left( \begin{array}{ccc}
- C^{a_2 \phi_2 \phi_3}_{1~1~1} &  C^{a_2 \phi_2 \phi_3}_{\bar{\alpha}~\alpha~1} &  C^{a_2 \phi_3 \phi_2}_{\bar{\alpha}~\alpha~1} \\ \\
C^{\phi_2 a_2 \phi_3}_{\bar{\alpha}~\alpha~1} & - C^{\phi_2 a_2 \phi_3}_{1~1~1}  &
C^{\phi_2 \phi_3 a_2 }_{\bar{\alpha}~\alpha~1} \\ \\
C^{\phi_3 a_2 \phi_2 }_{\bar{\alpha}~\alpha~1} &  C^{\phi_3 \phi_2 a_2 }_{\bar{\alpha}~\alpha~1} &
-C^{\phi_3 a_2 \phi_2}_{1~1~1}
\end{array}
\right)
\left(
\begin{array}{c}
T^{A_2}_{\bar{\alpha}} \\ \\ T^{\phi_2}_{\bar{\alpha}} \\ \\ T^{\phi_3}_{\bar{\alpha}}
\end{array}
\right)
= \left( \begin{array}{ccc}
- \dfrac{2}{3\sqrt{3}} &  - \dfrac{2}{15\sqrt{3}} &  -\dfrac{2\sqrt{2}}{3\sqrt{15}} \\ \\
- \dfrac{2}{15\sqrt{3}} & - \dfrac{2}{3\sqrt{3}}  &  \dfrac{2\sqrt{2}}{3\sqrt{15}} \\ \\
-\dfrac{2\sqrt{2}}{3\sqrt{15}} &  \dfrac{2\sqrt{2}}{3\sqrt{15}} & -\dfrac{2}{3\sqrt{3}}
\end{array}
\right)
\left(
\begin{array}{c}
T^{A_2}_{\bar{\alpha}} \\ \\ T^{\phi_2}_{\bar{\alpha}} \\ \\ T^{\phi_3}_{\bar{\alpha}}
\end{array}
\right),\nonumber
\ee
so that
\begin{equation}
T^{A_2}_{\bar{\alpha}} : T^{\phi_2}_{\bar{\alpha}} : T^{\phi_3}_{\bar{\alpha}}
= 1 : -1 : (-\sqrt{\frac{8}{5}}),\nonumber
\end{equation}
and $S \phi_2 \phi_3$ gives
\be
0 = \left( \begin{array}{ccc}
- C^{S \phi_2 \phi_3}_{1~1~1} &  C^{S \phi_2 \phi_3}_{\bar{\alpha}~\alpha~1} &  C^{S \phi_3 \phi_2}_{\bar{\alpha}~\alpha~1} \\ \\
C^{\phi_2 S \phi_3}_{\bar{\alpha}~\alpha~1} & - C^{\phi_2 S \phi_3}_{1~1~1}  &
C^{\phi_2 \phi_3 S}_{\bar{\alpha}~\alpha~1} \\ \\
C^{\phi_3 S \phi_2 }_{\bar{\alpha}~\alpha~1} &  C^{\phi_3 \phi_2 S }_{\bar{\alpha}~\alpha~1} &
-C^{\phi_3 S \phi_2}_{1~1~1}
\end{array}
\right)
\left(
\begin{array}{c}
T^{S}_{\bar{\alpha}} \\ \\ T^{\phi_2}_{\bar{\alpha}} \\ \\ T^{\phi_3}_{\bar{\alpha}}
\end{array}
\right)
= \left( \begin{array}{ccc}
- \dfrac{5}{6\sqrt{3}} &  \dfrac{1}{6\sqrt{3}} &  \dfrac{\sqrt{5}}{3\sqrt{6}} \\ \\
\dfrac{1}{6\sqrt{3}} & - \dfrac{5}{6\sqrt{3}}  &  \dfrac{\sqrt{5}}{3\sqrt{6}} \\ \\
\dfrac{\sqrt{5}}{3\sqrt{6}} & \dfrac{\sqrt{5}}{3\sqrt{6}} & - \dfrac{5}{6\sqrt{3}}
\end{array}
\right)
\left(
\begin{array}{c}
T^{S}_{\bar{\alpha}} \\ \\ T^{\phi_2}_{\bar{\alpha}} \\ \\ T^{\phi_3}_{\bar{\alpha}}
\end{array}
\right),\nonumber
\ee
so that
\begin{equation}
T^{S}_{\bar{\alpha}} : T^{\phi_2}_{\bar{\alpha}} : T^{\phi_3}_{\bar{\alpha}}
= 1 : 1 : \sqrt{\frac{8}{5}}.\nonumber
\end{equation}

Altogether, the Goldstone mode is
\begin{equation}
\overrightarrow{G}_{(3,2,-\frac{5}{6})}
=  a_2 \widehat{A}^{(3,2,-\frac{5}{6})}_{(24)}
- S \widehat{E}^{(3,2,-\frac{5}{6})}_{(24)}
- \phi_2 \widehat{\Phi}^{(3,2,-\frac{5}{6})}_{(24)}
- \sqrt{\frac{8}{5}} \phi_3 \widehat{\Phi}^{(3,2,-\frac{5}{6})}_{(75)}.
\end{equation}

\section{Goldstone modes for SSB of $G_{51}$: [(3,2,$\frac{1}{6}$) + c.c.]}

In studying the SSB of the flipped $SU(5)$,
the fields need to be decomposed into $G_{51}=(SU(5) \otimes U(1))^{Flipped}$
representations and use the CGCs accordingly. Under $G_{51}$, the following representations
\begin{eqnarray}
\widetilde{a}_1 &\equiv& \widehat{A}_{(1,0)},~~~
\widetilde{a}_2 \equiv \widehat{A}_{(24,0)},~~~
\widetilde{S} \equiv \widehat{E}_{(24,0)}, \nonumber\\
{\delta} &\equiv& \widehat{\Delta}_{(\overline{50},-2)},~~~
{\overline{\delta}} \equiv \widehat{\overline{\Delta}}_{(50,2)},\nonumber\\
\widetilde{\phi}_1 &\equiv& \widehat{\Phi}_{(1,0)},  ~~~
\widetilde{\phi}_2 \equiv \widehat{\Phi}_{(24,0)}, ~~~
\widetilde{\phi}_3 \equiv \widehat{\Phi}_{(75,0)}. \label{fsu5VEV}
\end{eqnarray}
contain the SM singlets whose VEVs will be denoted by the same symbols.
Here the subscripts are the representations under $G_{51}$ .
These VEVs  are related to those under $G_{422}$ and $SU(5)$ by
\be
\widetilde{S} = E =S,
~~~~~{\delta} = v_R =V_R,
~~~~~{\overline{\delta}} = \overline{v_R} =\overline{V_R}, \label{vev1}
\ee
\begin{equation}
\left(\begin{array}{c}
 \widetilde{a}_1\\ \\ \widetilde{a}_2
\end{array}\right)
= \left[ \begin{array}{cc}
\sqrt{\dfrac{2}{5}} & - \sqrt{\dfrac{3}{5}} \\
\sqrt{\dfrac{3}{5}} & \sqrt{\dfrac{2}{5}}
\end{array} \right]
\left(\begin{array}{c}
A_1 \\ \\ A_2
\end{array}\right)
= \left[ \begin{array}{cc}
-\dfrac{1}{5} & \dfrac{2\sqrt{6}}{5} \\
\dfrac{2\sqrt{6}}{5} & \dfrac{1}{5}
\end{array} \right]
\left(\begin{array}{c}
a_1 \\ \\ a_2
\end{array}\right),\label{vev2}
\end{equation}	
and
\begin{eqnarray}
\left(\begin{array}{c}
 \widetilde{\phi}_1 \\ \\ \widetilde{\phi}_2 \\ \\ \widetilde{\phi}_3
\end{array}\right)
&=& \left[ \begin{array}{ccc}
\sqrt{\dfrac{1}{10}} & \sqrt{\dfrac{3}{10}} & -\sqrt{\dfrac{3}{5}} \\
\sqrt{\dfrac{2}{5}} & - \sqrt{\dfrac{8}{15}} & - \sqrt{\dfrac{1}{15}} \\
\sqrt{\dfrac{1}{2}} & \sqrt{\dfrac{1}{6}} & \sqrt{\dfrac{1}{3}}
\end{array} \right]
\left(\begin{array}{c}
\Phi_1 \\ \\ \Phi_2 \\ \\ \Phi_3
\end{array}\right) \nonumber \\
&=& \left[ \begin{array}{ccc}
-\dfrac{1}{5} & -\dfrac{2}{5} & \dfrac{2\sqrt{5}}{5} \\
-\dfrac{2}{5} & \dfrac{13}{15} & \dfrac{2\sqrt{5}}{15} \\
\dfrac{2\sqrt{5}}{5} & \dfrac{2\sqrt{5}}{15} & \dfrac{1}{3}
\end{array} \right]
\left(\begin{array}{c}
\phi_1 \\ \\ \phi_2 \\ \\ \phi_3
\end{array}\right).\label{vev3}
\end{eqnarray}	

The fields contain to $\bar{\alpha}=(3,2,\frac{1}{6})$ are
$$\widehat{A}^{(3,2,\frac{1}{6})}_{(24,0)},
\widehat{E}^{(3,2,\frac{1}{6})}_{(24,0)},
\widehat{D}^{(3,2,\frac{1}{6})}_{(10,-6)},
\widehat{\Delta}^{(3,2,\frac{1}{6})}_{(\overline{50},-2)},
\widehat{\overline{\Delta}}^{(3,2,\frac{1}{6})}_{(\overline{45}, -2)},
\widehat{\Phi}^{(3,2,\frac{1}{6})}_{(24,0)},
\widehat{\Phi}^{(3,2,\frac{1}{6})}_{(75,0)},$$
they are related to the fields under $G_{422}$ and $SU(5)$ by
\begin{eqnarray}
&&\widehat{A}^{(3,2,\frac{1}{6})}_{(24,0)} = \widehat{A}^{(3,2,\frac{1}{6})}_{(6,2,2)}
= \widehat{A}^{(3,2,\frac{1}{6})}_{(10, 4)}, \notag \\
&&\widehat{E}^{(3,2,\frac{1}{6})}_{(24,0)} = \widehat{E}^{(3,2,\frac{1}{6})}_{(6,2,2)}
=\widehat{E}^{(3,2,\frac{1}{6})}_{(15,4)}, \notag \\
&&\widehat{D}^{(3,2,\frac{1}{6})}_{(10,-6)}= \widehat{D}^{(3,2,\frac{1}{6})}_{(15,2,2)}
=\widehat{D}^{(3,2,\frac{1}{6})}_{(10, -6)}, \notag \\
&&\widehat{\Delta}^{(3,2,\frac{1}{6})}_{(\overline{50},-2)} =
\widehat{\Delta}^{(3,2,\frac{1}{6})}_{(15,2,2)}
=\widehat{\Delta}^{(3,2,\frac{1}{6})}_{(\overline{15}, 6)}, \notag \\
&&\widehat{\overline{\Delta}}^{(3,2,\frac{1}{6})}_{(\overline{45}, -2)} =
\widehat{\overline{\Delta}}^{(3,2,\frac{1}{6})}_{(15,2,2)}
=\widehat{\overline{\Delta}}^{(3,2,\frac{1}{6})}_{(15,-6)}, \notag
\end{eqnarray}
\begin{equation}
	\left(\begin{array}{c}
		\Phi^{(3,2,\frac{1}{6})}_{(24,0)} \\ \\ \Phi^{(3,2, \frac{1}{6})}_{(75,0)}
	\end{array}\right)
	= \left[ \begin{array}{cc}
		\sqrt{\dfrac{2}{3}} & \sqrt{\dfrac{1}{3}} \\
		- \sqrt{\dfrac{1}{3}} & \sqrt{\dfrac{2}{3}}
	\end{array} \right]
	\left(\begin{array}{c}
		\Phi^{(3,2,\frac{1}{6})}_{(6,2,2)} \\ \\ \Phi^{(3,2,\frac{1}{6})}_{(10,2,2)}
	\end{array}\right)
	= \left[ \begin{array}{cc}
		\dfrac{2\sqrt{2}}{3} & \dfrac{1}{3} \\
		\dfrac{1}{3} & -\dfrac{2\sqrt{2}}{3}
	\end{array} \right]
	\left(\begin{array}{c}
		\Phi^{(3,2,\frac{1}{6})}_{(\overline{10},0)} \\ \\ \Phi^{(3,2,\frac{1}{6})}_{(\overline{40},0)}
	\end{array}\right).\label{vev4}
\end{equation}
Note that $\widehat{D}_{(10,-6)}$ and $\widehat{\overline{\Delta}}{(\overline{45}, -2)}$
contain no VEV, so the Goldstone mode can be denoted as
\begin{eqnarray}
	\overrightarrow{G}_{(3,2,\frac{1}{6})} &=&
	T^{\widetilde{a}_2}_{\bar{\alpha}} \widetilde{a}_2 \widehat{A}^{(3,2,\frac{1}{6})}_{(24,0)}
	+ T^{\widetilde{S}}_{\bar{\alpha}} \widetilde{S} \widehat{E}^{(3,2,\frac{1}{6})}_{(24,0)}
	+ T^{\delta}_{\bar{\alpha}} \delta \widehat{\Delta}^{(3,2,\frac{1}{6})}_{(\overline{50},-2)} \notag \\
	&+& T^{\overline{\delta}}_{\bar{\alpha}} \overline{\delta} \widehat{\overline{\Delta}}^{(3,2,\frac{1}{6})}_{(\overline{45}, -2)}
	+ T^{\widetilde{\phi}_2}_{\bar{\alpha}} \widetilde{\phi}_2 \widehat{\Phi}^{(3,2,\frac{1}{6})}_{(24,0)}
	+ T^{\widetilde{\phi}_3}_{\bar{\alpha}} \widetilde{\phi}_3 \widehat{\Phi}^{(3,2,\frac{1}{6})}_{(75,0)}.\nonumber
\end{eqnarray}
The needed CGCs can be tranformed from those under $G_{422}$ through (\ref{vev1},\ref{vev2},\ref{vev3},\ref{vev4}),
and are summarized in Table \ref{tabfsu5a},\ref{tabfsu5b}.
For  completeness, the full mass matrix for [(3,2,$\frac{1}{6}$) + c.c.] can be also transformed
through (\ref{vev1},\ref{vev2},\ref{vev3},\ref{vev4}) and is given in the Appendix, for the readers
to check the results of this Section.

1) According to (\ref{case3}), the coupling $\widetilde{\phi}_2^3$ gives
\be
0 = C^{\widetilde{\phi}_2 \widetilde{\phi}_2 \widetilde{\phi}_2}_{1~1~1} - 2 C^{\widetilde{\phi}_2 \widetilde{\phi}_2 \widetilde{\phi}_2}_{\bar{\alpha}~\alpha~1} = (- \frac{7}{9\sqrt{15}}) - 2*(- \frac{7}{18\sqrt{15}}),\nonumber
\ee
$\widetilde{\phi}_3^3$ gives
\be
0 = C^{\widetilde{\phi}_3 \widetilde{\phi}_3 \widetilde{\phi}_3}_{1~1~1} - 2 C^{\widetilde{\phi}_3 \widetilde{\phi}_3 \widetilde{\phi}_3}_{\bar{\alpha}~\alpha~1} = \frac{8}{9\sqrt{3}} - 2*\frac{4}{9\sqrt{3}},\nonumber
\ee
and $\widetilde{S}^3$ gives
\begin{eqnarray}
0 =C^{\widetilde{S}\widetilde{S}\widetilde{S}}_{111} - 2 C^{\widetilde{S}\widetilde{S}\widetilde{S}}_{\bar{\alpha} \alpha 1} = \sqrt{\frac{3}{5}} - 2 *\frac{\sqrt{3}}{2\sqrt{5}}.\nonumber
\end{eqnarray}

2) $\widetilde{\phi}_1, \widetilde{a}_1$ contain neither Goldstones, the couplings
$\widetilde{\phi}_2^2 \widetilde{\phi}_1$£¬$\widetilde{\phi}_3^2 \widetilde{\phi}_1$£¬
$\widetilde{a}_2^2 \widetilde{\phi}_1$£¬$\widetilde{\phi}_2^2 \widetilde{a}_1$£¬$\widetilde{\phi}_3^2 \widetilde{a}_1$
lead to only trivial results.
The coupling $\delta \overline{\delta} \widetilde{\phi}_2$, according to (\ref{case1}), gives
\begin{equation}
\frac{T^{\widetilde{\phi}_2}_{\bar{\alpha}}}{T^{\delta}_{\bar{\alpha}}} = \frac{C^{ \delta \widetilde{\phi}_2 \overline{\delta}}_{\bar{\alpha}\alpha1}}{C^{ \delta \widetilde{\phi}_2 \overline{\delta}}_{111}}
= (-\frac{1}{30}) / (- \frac{1}{5\sqrt{15}}) = \frac{\sqrt{15}}{6},\nonumber
\end{equation}
$\delta \overline{\delta} \widetilde{\phi}_3$ gives
\begin{equation}
\frac{T^{\widetilde{\phi}_3}_{\bar{\alpha}}}{T^{\delta}_{\bar{\alpha}}} = \frac{C^{\delta \widetilde{\phi}_3  \overline{\delta}}_{\bar{\alpha}\alpha1}}{C^{ \delta \widetilde{\phi}_3 \overline{\delta}}_{111}}
= (-\frac{\sqrt{2}}{15}) / (\frac{1}{5\sqrt{3}}) = - \frac{\sqrt{6}}{3},\nonumber
\end{equation}
$\delta \overline{\delta} \widetilde{a}_2$ gives
\begin{equation}
\frac{T^{\widetilde{a}_2}_{\bar{\alpha}}}{T^{\delta}_{\bar{\alpha}}} = \frac{C^{\delta \widetilde{a}_2  \overline{\delta}}_{\bar{\alpha}\alpha1}}{C^{\delta \widetilde{a}_2  \overline{\delta}}_{111}}
= (-\frac{1}{5}) / (-\frac{2\sqrt{3}}{5\sqrt{5}}) = \frac{\sqrt{15}}{6},\nonumber
\end{equation}
$\widetilde{a}_2 \widetilde{\phi}_2 \widetilde{a}_1$ gives
\begin{equation}
\frac{T^{\widetilde{a}_2}_{\bar{\alpha}}}{T^{\widetilde{\phi}_2}_{\bar{\alpha}}} = \frac{C^{\widetilde{a}_2 \widetilde{\phi}_2 \widetilde{a}_1}_{\bar{\alpha}~\alpha~1}}{C^{\widetilde{a}_2 \widetilde{\phi}_2 \widetilde{a}_1}_{1~1~1}} = (\sqrt{\frac{2}{5}}) / (\sqrt{\frac{2}{5}}) = 1,\nonumber
\end{equation}
$\widetilde{a}_2 \widetilde{\phi}_2 \widetilde{\phi}_1$ gives
\begin{equation}
\frac{T^{\widetilde{a}_2}_{\bar{\alpha}}}{T^{\widetilde{\phi}_2}_{\bar{\alpha}}} = \frac{C^{\widetilde{a}_2 \widetilde{\phi}_2 \widetilde{\phi}_1}_{\bar{\alpha}~\alpha~1}}{C^{\widetilde{a}_2 \widetilde{\phi}_2 \widetilde{\phi}_1}_{1~1~1}} = (\frac{2\sqrt{3}}{5\sqrt{5}}) / (\frac{2\sqrt{3}}{5\sqrt{5}})= 1,\nonumber
\end{equation}
$\widetilde{S} \widetilde{a}_2 \widetilde{a}_1$ gives
\begin{equation}
\frac{T^{\widetilde{S}}_{\bar{\alpha}}}{T^{\widetilde{a}_2}_{\bar{\alpha}}} = \frac{C^{\widetilde{S} \widetilde{a}_2 \widetilde{a}_1}_{\bar{\alpha}~\alpha~1}}{C^{\widetilde{S} \widetilde{a}_2 \widetilde{a}_1}_{1~1~1}}
= (\sqrt{\frac{2}{5}}) / (\sqrt{\frac{2}{5}}) = 1,\nonumber
\end{equation}
and $\widetilde{S} \widetilde{\phi}_2 \widetilde{\phi}_1$ gives
\begin{equation}
\frac{T^{\widetilde{S}}_{\bar{\alpha}}}{T^{\widetilde{\phi}_2}_{\bar{\alpha}}} = \frac{C^{\widetilde{S} \widetilde{\phi}_2 \widetilde{\phi}_1}_{\bar{\alpha}~\alpha~1}}
{C^{\widetilde{S} \widetilde{\phi}_2 \widetilde{\phi}_1}_{1~1~1}} = (\frac{\sqrt{3}}{2\sqrt{5}} ) / (\frac{\sqrt{3}}{2\sqrt{5}} ) = 1.\nonumber
\end{equation}

3) According to (\ref{case2}),
$\widetilde{\phi}_2 \widetilde{\phi}_3^2$ gives
\begin{eqnarray}
\frac{T^{\widetilde{\phi}_2}_{\bar{\alpha}}}{T^{\widetilde{\phi}_3}_{\bar{\alpha}}}
&=& \frac{2 C^{\widetilde{\phi}_2 \widetilde{\phi}_3 \widetilde{\phi}_3}_{\bar{\alpha}~\alpha~1}}{C^{\widetilde{\phi}_2 \widetilde{\phi}_3 \widetilde{\phi}_3}_{1~1~1}}
= 2* (\frac{\sqrt{2}}{9\sqrt{3}}) / (-\frac{8}{9\sqrt{15}})
= - \sqrt{\frac{5}{8}}, \nonumber\\
or:
\frac{T^{\widetilde{\phi}_2}_{\bar{\alpha}}}{T^{\widetilde{\phi}_3}_{\bar{\alpha}}}
&=& \frac{C^{\widetilde{\phi}_3 \widetilde{\phi}_3 \widetilde{\phi}_2}_{1~1~1} - C^{\widetilde{\phi}_3 \widetilde{\phi}_3 \widetilde{\phi}_2}_{\bar{\alpha}~\alpha~1}}{C^{\widetilde{\phi}_3 \widetilde{\phi}_2 \widetilde{\phi}_3}_{\bar{\alpha}~\alpha~1}}
= (\frac{-8}{9\sqrt{15}} - \frac{-11}{18\sqrt{15}}) / (\frac{\sqrt{2}}{9\sqrt{3}})
= - \sqrt{\frac{5}{8}},\nonumber
\end{eqnarray}
$\widetilde{\phi}_2^2 \widetilde{\phi}_3$ gives
\begin{eqnarray}
\frac{T^{\widetilde{\phi}_2}_{\bar{\alpha}}}{T^{\widetilde{\phi}_3}_{\bar{\alpha}}}
&=& \frac{C^{\widetilde{\phi}_2 \widetilde{\phi}_3 \widetilde{\phi}_2}_{\bar{\alpha}~\alpha~1}}{C^{\widetilde{\phi}_2 \widetilde{\phi}_2 \widetilde{\phi}_3}_{1~1~1} - C^{\widetilde{\phi}_2 \widetilde{\phi}_2 \widetilde{\phi}_3}_{\bar{\alpha}~\alpha~1}}
= (-\frac{\sqrt{5}}{9\sqrt{6}}) / (\frac{5}{18\sqrt{3}} - \frac{1}{18\sqrt{3}})
= - \sqrt{\frac{5}{8}}, \nonumber\\
or:
\frac{T^{\widetilde{\phi}_2}_{\bar{\alpha}}}{T^{\widetilde{\phi}_3}_{\bar{\alpha}}}
&=& \frac{C^{\widetilde{\phi}_3 \widetilde{\phi}_2 \widetilde{\phi}_2}_{1~1~1} }{2 C^{\widetilde{\phi}_3 \widetilde{\phi}_2 \widetilde{\phi}_2}_{\bar{\alpha}~\alpha~1}}
= (\frac{5}{18\sqrt{3}}) / 2*(- \frac{\sqrt{5}}{9\sqrt{6}})
= - \sqrt{\frac{5}{8}},\nonumber
\end{eqnarray}
$\widetilde{a}_2^2 \widetilde{\phi}_2$ gives
\begin{eqnarray}
\frac{T^{\widetilde{a}_2}_{\bar{\alpha}}}{T^{\widetilde{\phi}_2}_{\bar{\alpha}}}
&=& \frac{C^{\widetilde{a}_2 \widetilde{\phi}_2 \widetilde{a}_2}_{\bar{\alpha}~\alpha~1}}{C^{\widetilde{a}_2 \widetilde{a}_2 \widetilde{\phi}_2}_{1~1~1} - C^{\widetilde{a}_2 \widetilde{a}_2 \widetilde{\phi}_2}_{\bar{\alpha}~\alpha~1}}
= (- \frac{1}{3\sqrt{15}}) / (\frac{-2}{3\sqrt{15}} - \frac{-1}{3\sqrt{15}}) = 1, \nonumber\\
or:
\frac{T^{\widetilde{a}_2}_{\bar{\alpha}}}{T^{\widetilde{\phi}_2}_{\bar{\alpha}}}
&=& \frac{C^{\widetilde{\phi}_2 \widetilde{a}_2 \widetilde{a}_2}_{1~1~1}}{2 C^{\widetilde{\phi}_2 \widetilde{a}_2 \widetilde{a}_2 }_{\bar{\alpha}~\alpha~1}}
= (-\frac{2}{3\sqrt{15}}) / 2*(-\frac{1}{3\sqrt{15}}) = 1,\nonumber
\end{eqnarray}
$\widetilde{a}_2^2 \widetilde{\phi}_3$ gives
\begin{eqnarray}
\frac{T^{\widetilde{a}_2}_{\bar{\alpha}}}{T^{\widetilde{\phi}_3}_{\bar{\alpha}}}
&=& \frac{C^{\widetilde{a}_2 \widetilde{\phi}_3 \widetilde{a}_2}_{\bar{\alpha}~\alpha~1}}{C^{\widetilde{a}_2 \widetilde{a}_2 \widetilde{\phi}_3}_{1~1~1} - C^{\widetilde{a}_2 \widetilde{a}_2 \widetilde{\phi}_3}_{\bar{\alpha}~\alpha~1}}
= (- \frac{\sqrt{10}}{3\sqrt{3}}) / (\frac{5}{3\sqrt{3}} - \frac{1}{3\sqrt{3}}) = -\sqrt{\frac{5}{8}}, \nonumber\\
or:
\frac{T^{\widetilde{a}_2}_{\bar{\alpha}}}{T^{\widetilde{\phi}_3}_{\bar{\alpha}}}
&=& \frac{C^{\widetilde{\phi}_3 \widetilde{a}_2 \widetilde{a}_2}_{1~1~1}}{2 C^{\widetilde{\phi}_3 \widetilde{a}_2 \widetilde{a}_2 }_{\bar{\alpha}~\alpha~1}}
= (\frac{5}{3\sqrt{3}}) / 2*(-\frac{\sqrt{10}}{3\sqrt{3}}) = -\sqrt{\frac{5}{8}},\nonumber
\end{eqnarray}
$\widetilde{a}_2 \widetilde{\phi}_2^2$ gives
\begin{eqnarray}
\frac{T^{\widetilde{a}_2}_{\bar{\alpha}}}{T^{\widetilde{\phi}_2}_{\bar{\alpha}}} &=& \frac{2 C^{\widetilde{a}_2 \widetilde{\phi}_2 \widetilde{\phi}_2}_{\bar{\alpha}~\alpha~1}}{C^{\widetilde{a}_2 \widetilde{\phi}_2 \widetilde{\phi}_2}_{1~1~1}} = 2*(-\frac{4}{15\sqrt{15}}) /(- \frac{8}{15\sqrt{15}}) = 1, \nonumber\\
or:
\frac{T^{\widetilde{a}_2}_{\bar{\alpha}}}{T^{\widetilde{\phi}_2}_{\bar{\alpha}}} &=& \frac{C^{\widetilde{\phi}_2 \widetilde{\phi}_2 \widetilde{a}_2}_{1~1~1}
	- C^{\widetilde{\phi}_2 \widetilde{\phi}_2 \widetilde{a}_2}_{\bar{\alpha}~\alpha~1} }
{C^{\widetilde{\phi}_2 \widetilde{a}_2 \widetilde{\phi}_2}_{\bar{\alpha}~\alpha~1}}
= (- \frac{8}{15\sqrt{15}} + \frac{4}{15\sqrt{15}}) / (- \frac{4}{15\sqrt{15}}) = 1,\nonumber
\end{eqnarray}
$\widetilde{a}_2 \widetilde{\phi}_3^2$ gives
\begin{eqnarray}
\frac{T^{\widetilde{a}_2}_{\bar{\alpha}}}{T^{\widetilde{\phi}_3}_{\bar{\alpha}}} &=& \frac{2 C^{\widetilde{a}_2 \widetilde{\phi}_3 \widetilde{\phi}_3}_{\bar{\alpha}~\alpha~1}}{C^{\widetilde{a}_2 \widetilde{\phi}_3 \widetilde{\phi}_3}_{1~1~1}} = 2*(- \frac{4\sqrt{2}}{15\sqrt{3}}) /(\frac{32}{15\sqrt{15}}) = -\sqrt{\frac{5}{8}},\nonumber \\
or:
\frac{T^{\widetilde{a}_2}_{\bar{\alpha}}}{T^{\widetilde{\phi}_3}_{\bar{\alpha}}} &=& \frac{C^{\widetilde{\phi}_3 \widetilde{\phi}_3 \widetilde{a}_2}_{1~1~1}
	- C^{\widetilde{\phi}_3 \widetilde{\phi}_3 \widetilde{a}_2}_{\bar{\alpha}~\alpha~1} }
{C^{\widetilde{\phi}_3 \widetilde{a}_2 \widetilde{\phi}_3}_{\bar{\alpha}~\alpha~1}}
= (\frac{32}{15\sqrt{15}} - \frac{22}{15\sqrt{15}}) / (- \frac{4\sqrt{2}}{15\sqrt{3}}) = -\sqrt{\frac{5}{8}},\nonumber
\end{eqnarray}
$\widetilde{S} \widetilde{a}_2^2$ gives
\begin{eqnarray}
\frac{T^{\widetilde{S}}_{\bar{\alpha}}}{T^{\widetilde{a}_2}_{\bar{\alpha}}} &=& \frac{2 C^{\widetilde{S} \widetilde{a}_2 \widetilde{a}_2}_{\bar{\alpha}~\alpha~1}}{C^{\widetilde{S} \widetilde{a}_2 \widetilde{a}_2}_{1~1~1}}
= 2*(\frac{1}{2\sqrt{15}} ) / (\frac{1}{\sqrt{15}} ) = 1,\nonumber \\
or:
\frac{T^{\widetilde{S}}_{\bar{\alpha}}}{T^{\widetilde{a}_2}_{\bar{\alpha}}} &=&
\frac{C^{\widetilde{a}_2 \widetilde{a}_2 \widetilde{S}}_{1~1~1}
- C^{\widetilde{a}_2 \widetilde{a}_2 \widetilde{S}}_{\bar{\alpha}~\alpha~1}}
{C^{\widetilde{a}_2 \widetilde{S} \widetilde{a}_2}_{\bar{\alpha}~\alpha~1}}
= \left(\frac{1}{\sqrt{15}} - \frac{1}{2\sqrt{15}}
\right)/\left(\frac{1}{2\sqrt{15}}\right) = 1,\nonumber
\end{eqnarray}
$\widetilde{S} \widetilde{\phi}_2^2$ gives
\begin{eqnarray}
\frac{T^{\widetilde{S}}_{\bar{\alpha}}}{T^{\widetilde{\phi}_2}_{\bar{\alpha}}} &=& \frac{2 C^{\widetilde{S} \widetilde{\phi}_2 \widetilde{\phi}_2}_{\bar{\alpha}~\alpha~1}}
{C^{\widetilde{S} \widetilde{\phi}_2 \widetilde{\phi}_2}_{1~1~1}}
= 2*(\frac{1}{12\sqrt{15}}) / (\frac{1}{6\sqrt{15}}) = 1, \nonumber\\
or:
\frac{T^{\widetilde{S}}_{\bar{\alpha}}}{T^{\widetilde{\phi}_2}_{\bar{\alpha}}} &=&
\frac{C^{\widetilde{\phi}_2 \widetilde{\phi}_2 \widetilde{S}}_{1~1~1}
	- C^{\widetilde{\phi}_2 \widetilde{\phi}_2 \widetilde{S}}_{\bar{\alpha}~\alpha~1}}
	{C^{\widetilde{\phi}_2 \widetilde{S} \widetilde{\phi}_2}_{\bar{\alpha}~\alpha~1}}
= \left(\frac{1}{6\sqrt{15}} - \frac{1}{12\sqrt{15}}\right)/\left(\frac{1}{12\sqrt{15}}\right) = 1,\nonumber
\end{eqnarray}
$\widetilde{S} \widetilde{\phi}_3^2$ gives
\begin{eqnarray}
\frac{T^{\widetilde{S}}_{\bar{\alpha}}}{T^{\widetilde{\phi}_3}_{\bar{\alpha}}} &=& \frac{2 C^{\widetilde{S} \widetilde{\phi}_3 \widetilde{\phi}_3}_{\bar{\alpha}~\alpha~1}}
{C^{\widetilde{S} \widetilde{\phi}_3 \widetilde{\phi}_3}_{1~1~1}} = 2*(- \frac{1}{3\sqrt{6}}) / (\frac{4}{3\sqrt{15}}) = - \sqrt{\frac{5}{8}}, \nonumber\\
or:
\frac{T^{\widetilde{S}}_{\bar{\alpha}}}{T^{\widetilde{\phi}_3}_{\bar{\alpha}}} &=&
\frac{C^{\widetilde{\phi}_3 \widetilde{\phi}_3 \widetilde{S}}_{1~1~1}
- C^{\widetilde{\phi}_3 \widetilde{\phi}_3 \widetilde{S}}_{\bar{\alpha}~\alpha~1}}
{C^{\widetilde{\phi}_3 \widetilde{S} \widetilde{\phi}_3}_{\bar{\alpha}~\alpha~1}}
= \left(\frac{4}{3\sqrt{15}} - \frac{11}{12\sqrt{15}}\right)/\left(-\frac{1}{3\sqrt{6}}\right)
= - \sqrt{\frac{5}{8}}.\nonumber
\end{eqnarray}

4) According to (\ref{19}), $\widetilde{a}_2 \widetilde{\phi}_2 \widetilde{\phi}_3$ gives
\be
0 = \left( \begin{array}{ccc}
- C^{\widetilde{a}_2 \widetilde{\phi}_2 \widetilde{\phi}_3}_{1~1~1} &  C^{\widetilde{a}_2 \widetilde{\phi}_2 \widetilde{\phi}_3}_{\bar{\alpha}~\alpha~1} &  C^{\widetilde{a}_2 \widetilde{\phi}_3 \widetilde{\phi}_2}_{\bar{\alpha}~\alpha~1} \\ \\
C^{\widetilde{\phi}_2 \widetilde{a}_2 \widetilde{\phi}_3}_{\bar{\alpha}~\alpha~1} & - C^{\widetilde{\phi}_2 \widetilde{a}_2 \widetilde{\phi}_3}_{1~1~1}  &
C^{\widetilde{\phi}_2 \widetilde{\phi}_3 \widetilde{a}_2 }_{\bar{\alpha}~\alpha~1} \\ \\
C^{\widetilde{\phi}_3 \widetilde{a}_2 \widetilde{\phi}_2 }_{\bar{\alpha}~\alpha~1} &  C^{\widetilde{\phi}_3 \widetilde{\phi}_2 \widetilde{a}_2 }_{\bar{\alpha}~\alpha~1} &
-C^{\widetilde{\phi}_3 \widetilde{a}_2 \widetilde{\phi}_2}_{1~1~1}
\end{array}
\right)
\left(
\begin{array}{c}
T^{\widetilde{a}_2}_{\bar{\alpha}} \\ \\ T^{\widetilde{\phi}_2}_{\bar{\alpha}} \\ \\ T^{\widetilde{\phi}_3}_{\bar{\alpha}}
\end{array}
\right) = \left( \begin{array}{ccc}
\dfrac{2}{3\sqrt{3}} &  - \dfrac{2}{15\sqrt{3}} &  \dfrac{2\sqrt{2}}{3\sqrt{15}} \\ \\
- \dfrac{2}{15\sqrt{3}} & \dfrac{2}{3\sqrt{3}}  &  \dfrac{2\sqrt{2}}{3\sqrt{15}} \\ \\
\dfrac{2\sqrt{2}}{3\sqrt{15}} &  \dfrac{2\sqrt{2}}{3\sqrt{15}} & \dfrac{2}{3\sqrt{3}}
\end{array}
\right)
\left(
\begin{array}{c}
T^{\widetilde{a}_2}_{\bar{\alpha}} \\ \\ T^{\widetilde{\phi}_2}_{\bar{\alpha}} \\ \\ T^{\widetilde{\phi}_3}_{\bar{\alpha}}
\end{array}
\right),\nonumber
\ee
so that
\begin{equation}
T^{\widetilde{a}_2}_{\bar{\alpha}} : T^{\widetilde{\phi}_2}_{\bar{\alpha}} : T^{\widetilde{\phi}_3}_{\bar{\alpha}}
= 1 : 1 : (-\sqrt{\frac{8}{5}}),\nonumber
\end{equation}
and
$\widetilde{S} \widetilde{\phi}_2 \widetilde{\phi}_3$ gives
\be
0= \left( \begin{array}{ccc}
- C^{\widetilde{S} \widetilde{\phi}_2 \widetilde{\phi}_3}_{1~1~1} &  C^{\widetilde{S} \widetilde{\phi}_2 \widetilde{\phi}_3}_{\bar{\alpha}~\alpha~1} &  C^{\widetilde{S} \widetilde{\phi}_3 \widetilde{\phi}_2}_{\bar{\alpha}~\alpha~1} \\ \\
C^{\widetilde{\phi}_2 \widetilde{S} \widetilde{\phi}_3}_{\bar{\alpha}~\alpha~1} & - C^{\widetilde{\phi}_2 \widetilde{S} \widetilde{\phi}_3}_{1~1~1}  &
C^{\widetilde{\phi}_2 \widetilde{\phi}_3 \widetilde{S}}_{\bar{\alpha}~\alpha~1} \\ \\
C^{\widetilde{\phi}_3 \widetilde{S} \widetilde{\phi}_2 }_{\bar{\alpha}~\alpha~1} &  C^{\widetilde{\phi}_3 \widetilde{\phi}_2 \widetilde{S} }_{\bar{\alpha}~\alpha~1} &
-C^{\widetilde{\phi}_3 \widetilde{S} \widetilde{\phi}_2}_{1~1~1}
\end{array}
\right)
\left(
\begin{array}{c}
T^{\widetilde{S}}_{\bar{\alpha}} \\ \\ T^{\widetilde{\phi}_2}_{\bar{\alpha}} \\ \\ T^{\widetilde{\phi}_3}_{\bar{\alpha}}
\end{array}
\right) = \left( \begin{array}{ccc}
- \dfrac{5}{6\sqrt{3}} &  \dfrac{1}{6\sqrt{3}} & - \dfrac{\sqrt{5}}{3\sqrt{6}} \\ \\
\dfrac{1}{6\sqrt{3}} & - \dfrac{5}{6\sqrt{3}}  & - \dfrac{\sqrt{5}}{3\sqrt{6}} \\ \\
- \dfrac{\sqrt{5}}{3\sqrt{6}} & - \dfrac{\sqrt{5}}{3\sqrt{6}} & - \dfrac{5}{6\sqrt{3}}
\end{array}
\right)
\left(
\begin{array}{c}
T^{\widetilde{S}}_{\bar{\alpha}} \\ \\ T^{\widetilde{\phi}_2}_{\bar{\alpha}} \\ \\ T^{\widetilde{\phi}_3}_{\bar{\alpha}}
\end{array}
\right),\nonumber
\ee
so that
\begin{equation}
T^{\widetilde{S}}_{\bar{\alpha}} : T^{\widetilde{\phi}_2}_{\bar{\alpha}} : T^{\widetilde{\phi}_3}_{\bar{\alpha}}
= 1 : 1 : (- \sqrt{\frac{8}{5}}).\nonumber
\end{equation}

Altogether,  the Goldstino is consistently
\begin{eqnarray}
\overrightarrow{G}_{(3,2,\frac{1}{6})} =
\widetilde{a}_2 \widehat{A}^{(3,2,\frac{1}{6})}_{(24,0)}
+  \widetilde{S} \widehat{E}^{(3,2,\frac{1}{6})}_{(24,0)}
+ \frac{2\sqrt{15}}{5} \delta \widehat{\Delta}^{(3,2,\frac{1}{6})}_{(\overline{50},-2)}
+  \widetilde{\phi}_2 \widehat{\Phi}^{(3,2,\frac{1}{6})}_{(24,0)}
 -\sqrt{\frac{8}{5}} \widetilde{\phi}_3 \widehat{\Phi}^{(3,2,\frac{1}{6})}_{(75,0)}.
\end{eqnarray}

\section{Summary}

We have studied the Goldstone modes of SSB in the SUSY SO(10) model with general
renormalizable couplings.
VEVs and CGCs determine the contents of these Goldstone modes while
the parameters of the model are irrelevant.
Identities among the CGCs are examined to be in accord with the general conclusions in \cite{su2cgc}.

We thank Z.-Y. Chen and Z.-X. Ren for early collaborations.
DXZ also thank  Y.-X. Liu and D. Yang for helpful discussions.

\section*{Appendix: Mass matrix for  [(3,2,$\frac{1}{6}$) + c.c.].}

For the  SSB of $G_{422}$ and $SU(5)$,
the mass matrices for the Goldstone representations can be found in
\cite{fuku} and \cite{czy}, respectively.
Here we give the mass matrix for  [(3,2,$\frac{1}{6}$) + c.c.] using $G_{51}$
as the maximal subgroup of SO(10).
This matrix is transformed from \cite{fuku} through (\ref{vev1},\ref{vev2},\ref{vev3},\ref{vev4}).
\begin{minipage}{16cm}
	$\left[{\bf{(3,2}}, \frac{1}{6}) +c.c. \right]$\\[.2cm]
	{\bf c:}
	$\widehat{A}^{(3,2,\frac{1}{6})}_{(24,0)}$,
	$\widehat{E}^{(3,2,\frac{1}{6})}_{(24,0)}$,
	$\widehat{D}^{(3,2,\frac{1}{6})}_{(10, -6)}$,
	$\widehat{\Delta}^{(3,2,\frac{1}{6})}_{(\overline{50},-2)}$,
	$\widehat{\overline{\Delta}}^{(3,2,\frac{1}{6})}_{(\overline{45},-2)}$,
	$\widehat{\Phi}^{(3,2,\frac{1}{6})}_{(24,0)}$,
	$\widehat{\Phi}^{(3,2,\frac{1}{6})}_{(75,0)}$\\[.15cm]
	{\bf r:}
	$\widehat{A}^{(\overline{3},2,-\frac{1}{6})}_{(24,0)}$,
	$\widehat{E}^{(\overline{3},2,-\frac{1}{6})}_{(24,0)}$,
	$\widehat{D}^{(\overline{3},2,-\frac{1}{6})}_{(\overline{10}, 6)}$,
	$\widehat{\overline{\Delta}}^{(\overline{3},2,-\frac{1}{6})}_{(50,2)}$,
	$\widehat{\Delta}^{(\overline{3},2,-\frac{1}{6})}_{(45,2)}$,
	$\widehat{\Phi}^{(\overline{3},2,-\frac{1}{6})}_{(24,0)}$,
	$\widehat{\Phi}^{(\overline{3},2,-\frac{1}{6})}_{(75,0)}$\\[.2cm]
	\begin{equation}
	\left(
	\begin{array}{ccccccc}
	m_{11}
	& \sqrt{\frac{2}{5}} \widetilde{a}_1 \lambda _9+\frac{\widetilde{a}_2 \lambda _9}{2 \sqrt{15}}
& -\frac{i {\overline{\delta}} \lambda _{19}}{\sqrt{10}} & -\frac{{\overline{\delta}} \lambda _6}{5}
& 0 & m_{16} & m_{17} \\
	\sqrt{\frac{2}{5}} \widetilde{a}_1 \lambda _9+\frac{\widetilde{a}_2 \lambda _9}{2 \sqrt{15}}
& m_5+\frac{1}{2} \sqrt{\frac{3}{5}} \widetilde{S} \lambda _8 & 0 & 0 & -\frac{2 {\delta} \lambda_{12}}{5}
	& m_{26} & m_{27} \\
	\frac{i {\delta} \lambda _{18}}{\sqrt{10}} & 0 & m_{33} & m_{34} & m_{35}
& -\frac{{\delta} \lambda _{20}}{6 \sqrt{10}} & 	-\frac{{\delta} \lambda _{20}}{3 \sqrt{5}} \\
	-\frac{{\delta} \lambda_6}{5} & 0 & m_{43} & m_{44} & \frac{\widetilde{S} \lambda _{12}}{\sqrt{15}} & -\frac{{\delta} \lambda _2}{30} & -\frac{\sqrt{2}}{15}  {\delta} \lambda _2 \\
	0 & -\frac{2 {\delta}\lambda _{11}}{5} & m_{53} & \frac{\widetilde{S} \lambda _{11}}{\sqrt{15}} & m_{55} & 0 & 0 \\
	m_{16} & m_{26} & -\frac{\overline{{\delta}} \lambda _{21}}{6 \sqrt{10}} & -\frac{\overline{{\delta}} \lambda _2}{30} & 0 & m_{66} & m_{67} \\
	m_{17} & m_{27} & -\frac{\overline{{\delta}} \lambda _{21}}{3 \sqrt{5}} & -\frac{\sqrt{2}}{15}  \overline{{\delta}} \lambda _2 & 0
	& m_{67} & m_{77} \\
	\end{array}
	\right),\nonumber
	\end{equation}
\end{minipage}\\[.2cm]
where:
\bea
m_{11} &=& m_4+\frac{{\delta} \lambda _9}{2 \sqrt{15}}-\frac{\lambda _5\widetilde{\phi}_1}{\sqrt{15}}-\frac{\lambda _5 \widetilde{\phi}_2}{3 \sqrt{15}}+\frac{\lambda _5 \widetilde{\phi}_3}{3 \sqrt{3}}, \nonumber\\
m_{16} &=& \sqrt{\frac{2}{5}} \widetilde{a}_1 \lambda _5-\frac{\widetilde{a}_2 \lambda _5}{3 \sqrt{15}}+\frac{2}{5} \sqrt{\frac{3}{5}} \lambda _7 \widetilde{\phi}_1-\frac{4 \lambda
	_7 \widetilde{\phi}_2}{15 \sqrt{15}}-\frac{2 \lambda _7 \widetilde{\phi}_3}{15 \sqrt{3}},\nonumber\\
m_{17} &=& -\frac{1}{3} \sqrt{\frac{10}{3}} \widetilde{a}_2 \lambda _5+\frac{2}{3} \sqrt{\frac{2}{15}}
\lambda _7 \widetilde{\phi}_2-\frac{4}{15} \sqrt{\frac{2}{3}} \lambda _7 \widetilde{\phi}_3, \nonumber\\
m_{26} &=&\frac{1}{2} \sqrt{\frac{3}{5}} \lambda _{10} \widetilde{\phi}_1+\frac{\lambda _{10} \widetilde{\phi}_2}{12 \sqrt{15}}+\frac{\lambda _{10} \widetilde{\phi}_3}{6\sqrt{3}},
\nonumber\\
m_{27} &=& -\frac{1}{3} \sqrt{\frac{5}{6}} \lambda _{10} \widetilde{\phi}_2-\frac{\lambda _{10} \widetilde{\phi}_3}{3 \sqrt{6}},\nonumber\\
m_{33} &=& m_6-\frac{{\delta} \lambda _{14}}{3 \sqrt{15}}-\frac{\lambda_{15} \widetilde{\phi}_1}{3 \sqrt{15}}-\frac{2 \lambda_{15}\widetilde{\phi}_2}{3 \sqrt{15}}+\frac{\lambda _{15}\widetilde{\phi}_3}{3 \sqrt{3}},\nonumber\\
m_{34} &=& -\frac{i \widetilde{a}_2 \lambda _{18}}{2 \sqrt{6}}+\frac{\lambda _{20} \widetilde{\phi}_2}{12 \sqrt{6}}-\frac{1}{3}
\sqrt{\frac{2}{15}} \lambda _{20} \widetilde{\phi}_3, \nonumber\\
m_{35} &=& -\frac{1}{5} i \widetilde{a}_1 \lambda _{19}-\frac{i \widetilde{a}_2 \lambda _{19}}{10 \sqrt{6}}+\frac{\lambda _{21} \widetilde{\phi}_1}{10
	\sqrt{6}}-\frac{13 \lambda _{21} \widetilde{\phi}_2}{60 \sqrt{6}}-\frac{\lambda _{21} \widetilde{\phi}_3}{6 \sqrt{30}}, \nonumber\\
m_{43} &=& \frac{i \widetilde{a}_2 \lambda _{19}}{2 \sqrt{6}}+\frac{\lambda _{21} \widetilde{\phi}_2}{12 \sqrt{6}}-\frac{1}{3} \sqrt{\frac{2}{15}}
\lambda _{21} \widetilde{\phi}_3, \nonumber\\
m_{44} &=& m_2+\frac{\widetilde{a}_1 \lambda _6}{5 \sqrt{10}}-\frac{7 \widetilde{a}_2 \lambda _6}{10 \sqrt{15}}-\frac{\lambda _2 \widetilde{\phi}_1}{10 \sqrt{15}}-\frac{7
	\lambda _2 \widetilde{\phi}_2}{60 \sqrt{15}}+\frac{\lambda _2 \widetilde{\phi}_3}{15 \sqrt{3}}, \nonumber\\
m_{53} &=& \frac{1}{5} i \widetilde{a}_1 \lambda _{18}+\frac{i \widetilde{a}_2 \lambda _{18}}{10 \sqrt{6}}+\frac{\lambda _{20} \widetilde{\phi}_1}{10
	\sqrt{6}}-\frac{13 \lambda _{20} \widetilde{\phi}_2}{60 \sqrt{6}}-\frac{\lambda _{20} \widetilde{\phi}_3}{6 \sqrt{30}}, \nonumber\\
m_{55} &=& m_2-\frac{\widetilde{a}_1
	\lambda _6}{5 \sqrt{10}}+\frac{7 \widetilde{a}_2 \lambda _6}{10 \sqrt{15}}-\frac{\lambda _2 \widetilde{\phi}_2}{12 \sqrt{15}}+\frac{\lambda _2 \widetilde{\phi}_3}{30 \sqrt{3}}, \nonumber\\
m_{66} &=& m_1+\frac{4}{5} \sqrt{\frac{2}{5}}
\widetilde{a}_1 \lambda _7-\frac{4 \widetilde{a}_2 \lambda _7}{15 \sqrt{15}}+\frac{{\delta} \lambda _{10}}{12 \sqrt{15}}+\frac{\lambda _1 \widetilde{\phi}_1}{2 \sqrt{15}}-\frac{7 \lambda
	_1 \widetilde{\phi}_2}{18 \sqrt{15}}+\frac{\lambda _1 \widetilde{\phi}_3}{18 \sqrt{3}}, \nonumber\\
m_{67} &=& \frac{2}{3} \sqrt{\frac{2}{15}} \widetilde{a}_2 \lambda _7-\frac{1}{3} \sqrt{\frac{5}{6}}
\widetilde{S} \lambda _{10}-\frac{1}{9} \sqrt{\frac{5}{6}} \lambda _1 \widetilde{\phi}_2+\frac{1}{9} \sqrt{\frac{2}{3}} \lambda _1 \widetilde{\phi}_3, \nonumber\\
m_{77} &=& m_1-\frac{2}{5} \sqrt{\frac{2}{5}} \widetilde{a}_1 \lambda _7+\frac{22
	\widetilde{a}_2 \lambda _7}{15 \sqrt{15}}+\frac{11 {\delta} \lambda _{10}}{12 \sqrt{15}}-\frac{\lambda _1 \widetilde{\phi}_1}{\sqrt{15}}-\frac{11 \lambda _1 \widetilde{\phi}_2}{18 \sqrt{15}}+\frac{4
	\lambda _1 \widetilde{\phi}_3}{9 \sqrt{3}}.\nonumber
\eea

\newpage

\begin{table}[p]
\caption{The Goldstone modes and the corresponding SSB.}\label{tabsymm}
	\begin{center}
\begin{tabular}{|c||c|c|}
\hline\hline
  Goldstones & $G_1$ & $G_2$ \\
  \hline\hline
  $(1,1,0)$ & $U(1)_{I_{3R}}\otimes U(1)_{B-L}$ & $U(1)_Y$ \\
  $(1,1,1)+{\rm c.c.}$ & $SU(2)_R$&  $U(1)_{I_{3R}}$  \\
  $(3,1,\frac{2}{3})+{\rm c.c.}$ & $SU(4)_C$ & $SU(3)_C\otimes U(1)_{B-L}$ \\ \hline
  $(3,2, -\frac{5}{6})+{\rm c.c.}$ & $SU(5)$ & $G_{321}$ \\ \hline
  $(3,2,\frac{1}{6})+{\rm c.c.}$ & $G_{51}=(SU(5) \otimes U(1))^{Flipped}$
& $SU(3)_C \otimes SU(2)_L \otimes U(1)^\prime\otimes U(1)_X$ \\
  \hline\hline
\end{tabular}
\end{center}
\end{table}

\begin{table}[p]
	\caption{CGCs for the SM singlets (1,1,0) under $G_{422}$ SSB.
These SM singlets also correspond to Goldstones  (1,1,0) in the SSB of $U(1)_{I_{3R}}\otimes U(1)_{B-L}$ into $U(1)_Y$.
Hereafter, the CGCs for $A\Phi\Phi$ are understood as those for $\frac{1}{120}\epsilon A\Phi\Phi$\cite{fuku}.
Here, for example, ${A}_1$ stands for $\widehat{A}_{(1,1,3)}^{(1,1,0)}$.
For $X=\widehat{\overline{\Delta}}^{(1,1,0)}_{(10,1,3)}$
and $Y=\widehat{\Delta}^{(1,1,0)}_{(\overline{10},1,3)}$,  $XY{A}_1$
corresponds to the CGC $C^{\overline{v_{R}}v_RA_1}_{1 ~1 ~1}$.
}\label{tab110}
	\begin{center}
		\begin{tabular}{|cc|cc|c|ccc|}
			\hline\hline
			$X$ & $Y$ &
			$X Y {A}_1$ & $X Y {A}_2$ & $X Y {E}$ &
			$X Y {\Phi}_1 $ &
			$X Y {\Phi}_2 $ &
			$X Y {\Phi}_3 $
			\\
			\hline \hline
			$\widehat{A}^{(1,1,0)}_{(1,1,3)}$
			& $\widehat{A}^{(1,1,0)}_{(1,1,3)}$ & $0$ & $0$
			& $\frac{\sqrt{3}}{2\sqrt{5}}$
			& $\frac{1}{\sqrt{6}}$ & $0$ & $0$
			\\
			$\widehat{A}^{(1,1,0)}_{(15,1,1)}$ & $\widehat{A}^{(1,1,0)}_{(15,1,1)}$
			& $0$ & $0$ & $-\frac{1}{\sqrt{15}}$
			& $0$ & $\frac{\sqrt{2}}{3}$ & $0$
			\\
			$\widehat{A}^{(1,1,0)}_{(1,1,3)}$ & $\widehat{A}^{(1,1,0)}_{(15,1,1)}$
			& $0$ & $0$ & $0$
			& $0$ & $0$ & $\frac{1}{\sqrt{6}}$
			\\
			\hline
			$\widehat{E}^{(1,1,0)}_{(1,1,1)}$ & $\widehat{E}^{(1,1,0)}_{(1,1,1)}$
			& $0$ & $0$
			& $\frac{1}{2\sqrt{15}}$
			& $0$ & $0$ & $0$
			\\
			\hline
			$\widehat{\overline{\Delta}}^{(1,1,0)}_{(10,1,3)}$
			& $\widehat{\Delta}^{(1,1,0)}_{(\overline{10},1,3)}$
			& $-\frac{i}{5}$ & $-\frac{i\sqrt{3}}{5\sqrt{2}}$ & $0$
			& $\frac{1}{10\sqrt{6}}$ & $\frac{1}{10\sqrt{2}}$ & $\frac{1}{10}$
			\\
			\hline
			$\widehat{\Phi}^{(1,1,0)}_{(1,1,1)}$
			& $\widehat{\Phi}^{(1,1,0)}_{(1,1,1)}$
			& $0$ & $0$
			& $\frac{\sqrt{3}}{2\sqrt{5}}$
			& $0$ & $0$ & $0$
			\\
			$\widehat{\Phi}^{(1,1,0)}_{(1,1,1)}$
			& $\widehat{\Phi}^{(1,1,0)}_{(15,1,1)}$
			& $0$ & $\frac{\sqrt{2}}{5}$
			& $0$ & $0$ & $0$
			& $0$
			\\
			$\widehat{\Phi}^{(1,1,0)}_{(1,1,1)}$
			& $\widehat{\Phi}^{(1,1,0)}_{(15,1,3)}$
			& $0$ & $0$
			& $0$
			& $0$ & $0$ & $\frac{1}{6\sqrt{6}}$
			\\
			$\widehat{\Phi}^{(1,1,0)}_{(15,1,1)}$
			& $\widehat{\Phi}^{(1,1,0)}_{(15,1,1)}$
			& $0$ & $0$
			& $-\frac{1}{\sqrt{15}}$
			& $0$ & $\frac{1}{9\sqrt{2}}$ & $0$
			\\
			$\widehat{\Phi}^{(1,1,0)}_{(15,1,1)}$
			& $\widehat{\Phi}^{(1,1,0)}_{(15,1,3)}$
			& $\frac{\sqrt{2}}{5}$ & $0$
			& $0$
			& $0$ & $0$ & $\frac{1}{9\sqrt{2}}$
			\\
			$\widehat{\Phi}^{(1,1,0)}_{(15,1,3)}$
			& $\widehat{\Phi}^{(1,1,0)}_{(15,1,3)}$
			& $0$ & $\frac{2\sqrt{2}}{5\sqrt{3}}$
			& $\frac{1}{4\sqrt{15}}$
			& $\frac{1}{6\sqrt{6}}$ & $\frac{1}{9\sqrt{2}}$ & $0$
			\\
			\hline \hline
		\end{tabular}
	\end{center}
\end{table}

\begin{table}[p]
	\caption{CGCs for the Goldstones  (1,1,1)+c.c. of  $G_{422}$ SSB.
Here, for example, $XY{A}_1$ for $X=\widehat{\Delta}^{(1,1,1)}_{(\overline{10},1,3)}$
and $Y=\widehat{\overline{\Delta}}^{(1,1,-1)}_{(10,1,3)}$ corresponds to the CGC
$C^{\overline{v_{R}} v_R A_2}_{\bar{\alpha} ~\alpha ~1}$ where $\bar{\alpha}=(1,1,1)$ under $G_{321}$.
}\label{tab111}
	\begin{center}
		\begin{tabular}{|cc|cc|c|cc|ccc|}
			\hline\hline
			$X$ & $Y$ &
			$X Y {A}_1$ & $X Y {A}_2$ & $X Y {E}$ &
			$X Y {\Delta}$ & $X Y {\overline{\Delta}}$ &
			$X Y {\Phi}_1 $ &
			$X Y {\Phi}_2 $ &
			$X Y {\Phi}_3 $
			\\
			\hline \hline
			$\widehat{A}^{(1,1,1)}_{(1,1,3)}$ & $\widehat{A}^{(1,1,-1)}_{(1,1,3)}$
			& $0$ & $0$ & $\frac{\sqrt{3}}{2\sqrt{5}}$ & $0$ & $0$
			& $\frac{1}{\sqrt{6}}$ & $0$ & $0$
			\\
			$\widehat{A}^{(1,1,1)}_{(1,1,3)}$
			& $\widehat{\overline{\Delta}}^{(1,1,-1)}_{(10,1,3)}$
			& $0$ & $0$ & $0$
			& $-\frac{1}{5}$ & $0$ & $0$ & $0$  & $0$
			\\
			$\widehat{A}^{(1,1,1)}_{(1,1,3)}$
			& $\widehat{\Phi}^{(1,1,-1)}_{(15,1,3)}$
			& $0$ & $-\frac{1}{\sqrt{6}}$ & $0$ & $0$ & $0$
			& $0$ & $-\frac{\sqrt{2}}{5}$ & $0$
			\\
			\hline
			$\widehat{\Delta}^{(1,1,1)}_{(\overline{10},1,3)}$
			&  $\widehat{\overline{\Delta}}^{(1,1,-1)}_{(10,1,3)}$
			& $0$ & $-\frac{i\sqrt{3}}{5\sqrt{2}}$ & $0$ & $0$ & $0$
			& $\frac{1}{10\sqrt{6}}$ & $\frac{1}{10\sqrt{2}}$ & $0$
			\\
			$\widehat{\Delta}^{(1,1,1)}_{(\overline{10},1,3)}$
			&  $\widehat{A}^{(1,1,-1)}_{(1,1,3)}$
			& $0$ & $0$ & $0$  & $0$
			& $-\frac{1}{5}$ & $0$ & $0$& $0$
			\\
			$\widehat{\Delta}^{(1,1,1)}_{(\overline{10},1,3)}$
			&  $\widehat{\Phi}^{(1,1,-1)}_{(15,1,3)}$
			& $0$ & $0$ & $0$ & $0$
			& $-\frac{1}{10}$ & $0$ & $0$& $0$
			\\
			\hline
			$\widehat{\Phi}^{(1,1,1)}_{(15,1,3)}$
			& $\widehat{\Phi}^{(1,1,-1)}_{(15,1,3)}$
			& $0$ & $\frac{2\sqrt{2}}{5\sqrt{3}}$
			& $\frac{1}{4\sqrt{15}}$ & $0$ & $0$
			& $\frac{1}{6\sqrt{6}}$ & $\frac{1}{9\sqrt{2}}$ & $0$
			\\
			$\widehat{\Phi}^{(1,1,1)}_{(15,1,3)}$
			& $\widehat{\overline{\Delta}}^{(1,1,-1)}_{(10,1,3)}$
			& $0$ & $0$& $0$
			& $-\frac{1}{10}$ & $0$ & $0$ & $0$ & $0$
			\\
			\hline \hline
		\end{tabular}
	\end{center}
\end{table}

\begin{table}[p]
	\caption{CGCs for the Goldstones (3,1,$\frac{2}{3}$)+c.c. of  $G_{422}$ SSB.}\label{tab31}
	\begin{center}
		\begin{tabular}{|cc|cc|c|cc|ccc|}
			\hline\hline
			$X$ & $Y$ &
			$X Y {A}_1$ & $X Y {A}_2$ & $X Y {E}$ &
			$X Y {\Delta}$ & $X Y {\overline{\Delta}}$ &
			$X Y {\Phi}_1 $ &
			$X Y {\Phi}_2 $ &
			$X Y {\Phi}_3 $
			\\
			\hline \hline
			$\widehat{A}^{(3,1,\frac{2}{3})}_{(15,1,1)}$ & $\widehat{A}^{(\overline{3},1,-\frac{2}{3})}_{(15,1,1)}$
			& $0$ & $0$
			& $-\frac{1}{\sqrt{15}}$ & $0$ & $0$
			& $0$ & $\frac{1}{3\sqrt{2}}$ & $0$
			\\
			$\widehat{A}^{(3,1,\frac{2}{3})}_{(15,1,1)}$
			& $\widehat{\Delta}^{(\overline{3},1,-\frac{2}{3})}_{(\overline{10},1,3)}$
			& $0$ & $0$ & $0$ & $0$
			& $-\frac{1}{5}$ & $0$ & $0$& $0$
			\\
			$\widehat{A}^{(3,1,\frac{2}{3})}_{(15,1,1)}$
			& $\widehat{\Phi}^{(\overline{3},1,-\frac{2}{3})}_{(15,1,1)}$
			& $0$ & $-\frac{1}{3\sqrt{2}}$ & $0$ & $0$ & $0$
			& $-\frac{\sqrt{2}}{5}$ & $0$ & $0$
			\\
			$\widehat{A}^{(\overline{3},1,-\frac{2}{3})}_{(15,1,1)}$
			& $\widehat{\Phi}^{(3,1,\frac{2}{3})}_{(15,1,3)}$
			& $-\frac{1}{\sqrt{6}}$ & $0$  & $0$ & $0$ & $0$
			& $0$ & $0$ & $-\frac{\sqrt{2}}{5\sqrt{3}}$
			\\
			\hline
			$\widehat{\overline{\Delta}}^{(3,1,\frac{2}{3})}_{(10,1,3)}$
			& $\widehat{\Delta}^{(\overline{3},1,-\frac{2}{3})}_{(\overline{10},1,3)}$
			& $-\frac{i}{5}$ & $-\frac{i}{5\sqrt{6}}$ & $0$ & $0$ & $0$
			& $\frac{1}{10\sqrt{6}}$ & $\frac{1}{30\sqrt{2}}$ & $\frac{1}{30}$
			\\
			$\widehat{\overline{\Delta}}^{(3,1,\frac{2}{3})}_{(10,1,3)}$
			&  $\widehat{A}^{(\overline{3},1,-\frac{2}{3})}_{(15,1,1)}$
			& $0$ & $0$ & $0$
			& $-\frac{1}{5}$ & $0$ & $0$ & $0$& $0$
			\\
			$\widehat{\overline{\Delta}}^{(3,1,\frac{2}{3})}_{(10,1,3)}$
			&  $\widehat{\Phi}^{(\overline{3},1,-\frac{2}{3})}_{(15,1,3)}$
			& $0$ & $0$& $0$
			& $-\frac{1}{5\sqrt{6}}$ & $0$ & $0$ & $0$ & $0$
			\\
			\hline
			$\widehat{\Phi}^{(3,1,\frac{2}{3})}_{(15,1,1)}$
			& $\widehat{\Phi}^{(\overline{3},1,-\frac{2}{3})}_{(15,1,1)}$
			& $0$ & $0$
			& $-\frac{1}{\sqrt{15}}$ & $0$ & $0$
			& $0$ & $\frac{1}{18\sqrt{2}}$ & $0$
			\\
			$\widehat{\Phi}^{(3,1,\frac{2}{3})}_{(15,1,1)}$
			& $\widehat{\Phi}^{(\overline{3},1,-\frac{2}{3})}_{(15,1,3)}$
			& $\frac{\sqrt{2}}{5}$ & $0$
			& $0$ & $0$ & $0$
			& $0$ & $0$ & $\frac{1}{18\sqrt{2}}$
			\\
			$\widehat{\Phi}^{(3,1,\frac{2}{3})}_{(15,1,3)}$
			& $\widehat{\Phi}^{(\overline{3},1,-\frac{2}{3})}_{(15,1,3)}$
			& $0$ & $\frac{\sqrt{2}}{5\sqrt{3}}$
			& $\frac{1}{4\sqrt{15}}$ & $0$ & $0$
			& $\frac{1}{6\sqrt{6}}$ & $\frac{1}{18\sqrt{2}}$ & $0$
			\\
			$\widehat{\Phi}^{(\overline{3},1,-\frac{2}{3})}_{(15,1,1)}$
			& $\widehat{\overline{\Delta}}^{(3,1,\frac{2}{3})}_{(10,1,3)}$
			& $0$ & $0$ & $0$
			& $-\frac{1}{10\sqrt{3}}$ & $0$ & $0$& $0$ & $0$
			\\
			$\widehat{\Phi}^{(\overline{3},1,-\frac{2}{3})}_{(15,1,3)}$
			& $\widehat{\overline{\Delta}}^{(3,1,\frac{2}{3})}_{(10,1,3)}$
			& $0$ & $0$ & $0$
			& $-\frac{1}{5\sqrt{6}}$ & $0$ & $0$& $0$ & $0$
			\\
			\hline \hline
		\end{tabular}
	\end{center}
\end{table}

\begin{table}[p]
	\caption{CGCs for the SM singlets (1,1,0) under $SU(5)$.
Here, for example, $a_2$ stands for  $\widehat{A}_{(24)}^{(1,1,0)}$ {\it etc.}.}\label{tabsu5a}
	\begin{center}
		\begin{tabular}{|cc|cc|c|ccc|}
			\hline\hline
			$X$ & $Y$ &
			$X Y a_1$ & $X Y a_2$ & $X Y S$ &
			$X Y {\phi}_1 $ &
			$X Y {\phi}_2 $ &
			$X Y {\phi}_3 $
			\\
			\hline \hline
			$\widehat{A}^{(1,1,0)}_{(1)}$
			& $\widehat{A}^{(1,1,0)}_{(1)}$ & $0$ & $0$
			& $0$
			& $\frac{4}{\sqrt{15}}$ & $0$ & $0$
			\\
			$\widehat{A}^{(1,1,0)}_{(24)}$ & $\widehat{A}^{(1,1,0)}_{(24)}$
			& $0$ & $0$ & $\frac{1}{\sqrt{15}}$
			& $-\frac{1}{\sqrt{15}}$ & $-\frac{2}{3\sqrt{15}}$ & $\frac{5}{3\sqrt{3}}$
			\\
			$\widehat{A}^{(1,1,0)}_{(1)}$ & $\widehat{A}^{(1,1,0)}_{(24)}$
			& $0$ & $0$ & $\sqrt{\frac{2}{5}}$
			& $0$ & $\sqrt{\frac{2}{5}}$ & $0$
			\\
			\hline
			$\widehat{E}^{(1,1,0)}_{(24)}$ & $\widehat{E}^{(1,1,0)}_{(24)}$
			& $0$ & $0$
			& $\sqrt{\frac{3}{5}}$
			& $0$ & $0$ & $0$
			\\
			\hline
			$\widehat{\overline{\Delta}}^{(1,1,0)}_{(1)}$
			& $\widehat{\Delta}^{(1,1,0)}_{(1)}$
			& $-\frac{i}{\sqrt{10}}$ & $0$ & $0$
			& $\frac{1}{2\sqrt{15}}$ & $0$ & $0$
			\\
			\hline
			$\widehat{\Phi}^{(1,1,0)}_{(1)}$
			& $\widehat{\Phi}^{(1,1,0)}_{(1)}$
			& $\frac{6\sqrt{2}}{5\sqrt{5}}$ & $0$
			& $0$
			& $\sqrt{\frac{3}{5}}$ & $0$ & $0$
			\\
			$\widehat{\Phi}^{(1,1,0)}_{(1)}$
			& $\widehat{\Phi}^{(1,1,0)}_{(24)}$
			& $0$ & $-\frac{2\sqrt{3}}{5\sqrt{5}}$
			& $\frac{\sqrt{3}}{2\sqrt{5}}$ & $0$ & $\frac{1}{2\sqrt{15}}$
			& $0$
			\\
			$\widehat{\Phi}^{(1,1,0)}_{(1)}$
			& $\widehat{\Phi}^{(1,1,0)}_{(75)}$
			& $0$ & $0$
			& $0$
			& $0$ & $0$ & $-\frac{1}{\sqrt{15}}$
			\\
			$\widehat{\Phi}^{(1,1,0)}_{(24)}$
			& $\widehat{\Phi}^{(1,1,0)}_{(24)}$
			& $-\frac{4\sqrt{2}}{5\sqrt{5}}$ & $\frac{8}{15\sqrt{15}}$
			& $\frac{1}{6\sqrt{15}}$
			& $\frac{1}{2\sqrt{15}}$ & $-\frac{7}{9\sqrt{15}}$ & $\frac{5}{18\sqrt{3}}$
			\\
			$\widehat{\Phi}^{(1,1,0)}_{(24)}$
			& $\widehat{\Phi}^{(1,1,0)}_{(75)}$
			& $0$ & $\frac{2}{3\sqrt{3}}$
			& $\frac{5}{6\sqrt{3}}$
			& $0$ & $\frac{5}{18\sqrt{3}}$ & $-\frac{8}{9\sqrt{15}}$
			\\
			$\widehat{\Phi}^{(1,1,0)}_{(75)}$
			& $\widehat{\Phi}^{(1,1,0)}_{(75)}$
			& $\frac{2\sqrt{2}}{5\sqrt{5}}$ & $-\frac{32}{15\sqrt{15}}$
			& $\frac{4}{3\sqrt{15}}$
			& $-\frac{1}{\sqrt{15}}$ & $-\frac{8}{9\sqrt{15}}$ & $\frac{8}{9\sqrt{3}}$
			\\
			\hline \hline
		\end{tabular}
	\end{center}
\end{table}

\begin{table}[p]
	\caption{CGCs for the Goldstones (3,2,$- \frac{5}{6}$)+c.c. of $SU(5)$ SSB.}\label{tabsu5b}
	\begin{center}
		\begin{tabular}{|cc|cc|c|ccc|}
			\hline\hline
			$X$ & $Y$ &
			$X Y a_1$ & $X Y a_2$ & $X Y S$ &
			$X Y {\phi}_1 $ &
			$X Y {\phi}_2 $ &
			$X Y {\phi}_3 $
			\\
			\hline \hline
			$\widehat{A}^{(3,2,-\frac{5}{6})}_{(24)}$ & $\widehat{A}^{(\overline{3},2,\frac{5}{6})}_{(24)}$
			& $0$ & $0$
			& $\frac{1}{2\sqrt{15}}$
			& $- \frac{1}{\sqrt{15}}$ & $- \frac{1}{3\sqrt{15}}$ & $\frac{1}{3 \sqrt{3}}$
			\\
			$\widehat{A}^{(3,2,-\frac{5}{6})}_{(24)}$
			&  $\widehat{E}^{(\overline{3},2,\frac{5}{6})}_{(24)}$
			& $-\sqrt{\frac{2}{5}}$ & $-\frac{1}{2\sqrt{15}}$
			& $0$ & $0$ & $0$ & $0$
			\\
			$\widehat{A}^{(3,2,-\frac{5}{6})}_{(24)}$
			& $\widehat{\Phi}^{(\overline{3},2,\frac{5}{6})}_{(24)}$
			& $-\sqrt{\frac{2}{5}}$ & $\frac{1}{3\sqrt{15}}$ & $0$ & $\frac{2\sqrt{3}}{5\sqrt{5}}$
			 & $- \frac{4}{15\sqrt{15}}$ & $- \frac{2}{15\sqrt{3}}$
			\\
			$\widehat{A}^{(3,2,-\frac{5}{6})}_{(24)}$
			& $\widehat{\Phi}^{(\overline{3},2,\frac{5}{6})}_{(75)}$
			& $0$ & $-\frac{\sqrt{10}}{3\sqrt{3}}$ & $0$
			& $0$ & $-\frac{2\sqrt{2}}{3\sqrt{15}}$ & $\frac{4\sqrt{2}}{15\sqrt{3}}$
			\\
			\hline
			$\widehat{E}^{(3,2,-\frac{5}{6})}_{(24)}$
			& $\widehat{E}^{(\overline{3},2,\frac{5}{6})}_{(24)}$
			& $0$ & $0$
			& $\frac{\sqrt{3}}{2\sqrt{5}}$ & $0$ & $0$ & $0$
			\\	
			$\widehat{E}^{(3,2,-\frac{5}{6})}_{(24)}$
			& $\widehat{\Phi}^{(\overline{3},2,\frac{5}{6})}_{(24)}$
			& $0$ & $0$ & $0$
			& $\frac{\sqrt{3}}{2\sqrt{5}}$ & $\frac{1}{12\sqrt{15}}$ & $\frac{1}{6\sqrt{3}}$
			\\
			$\widehat{E}^{(3,2,-\frac{5}{6})}_{(24)}$
			& $\widehat{\Phi}^{(\overline{3},2,\frac{5}{6})}_{(75)}$
			& $0$ & $0$ & $0$
			& $0$ & $\frac{\sqrt{5}}{3\sqrt{6}}$ & $\frac{1}{3\sqrt{6}}$
			\\
			\hline
			$\widehat{\Phi}^{(3,2,-\frac{5}{6})}_{(24)}$
			& $\widehat{\Phi}^{(\overline{3},2,\frac{5}{6})}_{(24)}$
			& $-\frac{4\sqrt{2}}{5\sqrt{5}}$ & $\frac{4}{15\sqrt{15}}$
			& $\frac{1}{12\sqrt{15}}$
			& $\frac{1}{2\sqrt{15}}$ & $-\frac{7}{18\sqrt{15}}$ & $\frac{1}{18\sqrt{3}}$
			\\
			$\widehat{\Phi}^{(3,2,-\frac{5}{6})}_{(24)}$
			&  $\widehat{\Phi}^{(\overline{3},2,\frac{5}{6})}_{(75)}$
			& $0$ & $\frac{2\sqrt{2}}{3\sqrt{15}}$
			& $\frac{\sqrt{5}}{3\sqrt{6}}$
			& $0$ & $\frac{\sqrt{5}}{9\sqrt{6}}$ & $-\frac{\sqrt{2}}{9\sqrt{3}}$
			\\
			$\widehat{\Phi}^{(3,2,-\frac{5}{6})}_{(75)}$
			& $\widehat{\Phi}^{(\overline{3},2,\frac{5}{6})}_{(75)}$
			& $\frac{2\sqrt{2}}{5\sqrt{5}}$ & $-\frac{22}{15\sqrt{15}}$
			& $\frac{11}{12\sqrt{15}}$
			& $-\frac{1}{\sqrt{15}}$ & $-\frac{11}{18\sqrt{15}}$ & $\frac{4}{9\sqrt{3}}$
			\\
			\hline \hline
		\end{tabular}
	\end{center}
\end{table}

\begin{table}[p]
	\caption{CGCs for the SM singlets (1,1,0) through $G_{51}$.
Here, for example, $\widetilde{\phi}_3 $ stands for $\widehat{\Phi}_{(75,0)}^{(1,1,0)}$.}.
\label{tabfsu5a}
	\begin{center}
		\begin{tabular}{|cc|cc|c|ccc|}
\hline \hline
			$X$ & $Y$ &
			$X Y \widetilde{a}_1$ & $X Y \widetilde{a}_2$ & $X Y \widetilde{S}$ &
			$X Y \widetilde{\phi}_1 $ &
			$X Y \widetilde{\phi}_2 $ &
			$X Y \widetilde{\phi}_3 $
			\\
			\hline \hline
			$\widehat{A}^{(1,1,0)}_{(1,0)}$
			& $\widehat{A}^{(1,1,0)}_{(1,0)}$ & $0$ & $0$
			& $0$
			& $\frac{4}{\sqrt{15}}$ & $0$ & $0$
			\\
			$\widehat{A}^{(1,1,0)}_{(24,0)}$ & $\widehat{A}^{(1,1,0)}_{(24,0)}$
			& $0$ & $0$ & $\frac{1}{\sqrt{15}}$
			& $-\frac{1}{\sqrt{15}}$ & $-\frac{2}{3\sqrt{15}}$ & $\frac{5}{3\sqrt{3}}$
			\\
			$\widehat{A}^{(1,1,0)}_{(1,0)}$ & $\widehat{A}^{(1,1,0)}_{(24,0)}$
			& $0$ & $0$ & $\sqrt{\frac{2}{5}}$
			& $0$ & $\sqrt{\frac{2}{5}}$ & $0$
			\\
			\hline
			$\widehat{E}^{(1,1,0)}_{(24,0)}$ & $\widehat{E}^{(1,1,0)}_{(24,0)}$
			& $0$ & $0$
			& $\sqrt{\frac{3}{5}}$
			& $0$ & $0$ & $0$
			\\
			\hline
			$\widehat{\overline{\Delta}}^{(1,1,0)}_{(\bar{50},-2)}$
			& $\widehat{\Delta}^{(1,1,0)}_{(50,2)}$
			& $\frac{i}{5\sqrt{10}}$ & $-\frac{i2\sqrt{3}}{5\sqrt{5}}$ & $0$
			& $-\frac{1}{10\sqrt{15}}$ & $-\frac{1}{5\sqrt{15}}$ & $\frac{1}{5\sqrt{3}}$
			\\
			\hline
			$\widehat{\Phi}^{(1,1,0)}_{(1,0)}$
			& $\widehat{\Phi}^{(1,1,0)}_{(1,0)}$
			& $-\frac{6\sqrt{2}}{5\sqrt{5}}$ & $0$
			& $0$
			& $\sqrt{\frac{3}{5}}$ & $0$ & $0$
			\\
			$\widehat{\Phi}^{(1,1,0)}_{(1,0)}$
			& $\widehat{\Phi}^{(1,1,0)}_{(24,0)}$
			& $0$ & $\frac{2\sqrt{3}}{5\sqrt{5}}$
			& $\frac{\sqrt{3}}{2\sqrt{5}}$  & $0$ & $\frac{1}{2\sqrt{15}}$
			& $0$
			\\
			$\widehat{\Phi}^{(1,1,0)}_{(1,0)}$
			& $\widehat{\Phi}^{(1,1,0)}_{(75,0)}$
			& $0$ & $0$
			& $0$
			& $0$ & $0$ & $-\frac{1}{\sqrt{15}}$
			\\
			$\widehat{\Phi}^{(1,1,0)}_{(24,0)}$
			& $\widehat{\Phi}^{(1,1,0)}_{(24,0)}$
			& $\frac{4\sqrt{2}}{5\sqrt{5}}$ & $-\frac{8}{15\sqrt{15}}$
			& $\frac{1}{6\sqrt{15}}$
			& $\frac{1}{2\sqrt{15}}$ & $-\frac{7}{9\sqrt{15}}$ & $\frac{5}{18\sqrt{3}}$
			\\
			$\widehat{\Phi}^{(1,1,0)}_{(24,0)}$
			& $\widehat{\Phi}^{(1,1,0)}_{(75,0)}$
			& $0$ & $-\frac{2}{3\sqrt{3}}$
			& $\frac{5}{6\sqrt{3}}$
			& $0$ & $\frac{5}{18\sqrt{3}}$ & $-\frac{8}{9\sqrt{15}}$
			\\
			$\widehat{\Phi}^{(1,1,0)}_{(75,0)}$
			& $\widehat{\Phi}^{(1,1,0)}_{(75,0)}$
			& $-\frac{2\sqrt{2}}{5\sqrt{5}}$ & $\frac{32}{15\sqrt{15}}$
			& $\frac{4}{3\sqrt{15}}$
			& $-\frac{1}{15}$ & $-\frac{8}{9\sqrt{15}}$ & $\frac{8}{9\sqrt{3}}$
			\\
			\hline \hline
		\end{tabular}
	\end{center}
\end{table}

\begin{table}[p]
	\caption{CGCs for the Goldstone (3,2,$\frac{1}{6}$)+c.c. of  $G_{51}$ SSB.}\label{tabfsu5b}
	\begin{center}
		\begin{tabular}{|cc|cc|c|cc|ccc|}
			\hline\hline
			$X$ & $Y$ &
			$X Y \widetilde{a}_1$ & $X Y \widetilde{a}_2$ & $X Y \widetilde{S}$ &
            $X Y \delta$ & $X Y \overline{\delta}$ &
			$X Y \widetilde{\phi}_1 $ &
			$X Y \widetilde{\phi}_2 $ &
			$X Y \widetilde{\phi}_3 $
			\\
			\hline \hline
			$\widehat{A}^{(3,2,\frac{1}{6})}_{(24,0)}$ & $\widehat{A}^{(\overline{3},2,-\frac{1}{6})}_{(24,0)}$
			& $0$ & $0$
			& $\frac{1}{2\sqrt{15}}$ & $0$ & $0$
			& $-\frac{1}{\sqrt{15}}$ & $-\frac{1}{3\sqrt{15}}$ & $\frac{1}{3\sqrt{3}}$
			\\
			$\widehat{A}^{(3,2,\frac{1}{6})}_{(24,0)}$
			&  $\widehat{E}^{(\overline{3},2,-\frac{1}{6})}_{(24,0)}$
			& $\sqrt{\frac{2}{5}}$ & $\frac{1}{2\sqrt{15}}$
			& $0$ & $0$ & $0$ & $0$ & $0$ & $0$
			\\
			$\widehat{A}^{(3,2,\frac{1}{6})}_{(24,0)}$
			& $\widehat{\overline{\Delta}}^{(\overline{3},2,-\frac{1}{6})}_{(50,2)}$
			& $0$ & $0$ & $0$ & $-\frac{1}{5}$ & $0$ & $0$ & $0$ & $0$
			\\
			$\widehat{A}^{(3,2,\frac{1}{6})}_{(24,0)}$
			& $\widehat{\Phi}^{(\overline{3},2,-\frac{1}{6})}_{(24,0)}$
			& $\sqrt{\frac{2}{5}}$ & $-\frac{1}{3\sqrt{15}}$ & $0$
			& $0$
			& $0$ & $\frac{2\sqrt{3}}{5\sqrt{5}}$ & $-\frac{4}{15\sqrt{15}}$ & $-\frac{2}{15\sqrt{3}}$
			\\
			$\widehat{A}^{(3,2,\frac{1}{6})}_{(24,0)}$
			& $\widehat{\Phi}^{(\overline{3},2,-\frac{1}{6})}_{(75,0)}$
			& $0$ & $-\frac{\sqrt{10}}{3\sqrt{3}}$ & $0$
			& $0$
			& $0$ & $0$ & $\frac{2\sqrt{2}}{3\sqrt{15}}$ & $-\frac{4\sqrt{2}}{15\sqrt{3}}$
			\\
			\hline
			$\widehat{E}^{(3,2,\frac{1}{6})}_{(24,0)}$
			& $\widehat{E}^{(\overline{3},2,-\frac{1}{6})}_{(24,0)}$
			& $0$ & $0$
			& $\frac{\sqrt{3}}{2\sqrt{5}}$
			& $0$ & $0$ & $0$ & $0$ & $0$
			\\
			$\widehat{E}^{(3,2,\frac{1}{6})}_{(24,0)}$
			& $\widehat{\Phi}^{(\overline{3},2,-\frac{1}{6})}_{(24,0)}$
			& $0$ & $0$ & $0$ & $0$ & $0$
			& $\frac{\sqrt{3}}{2\sqrt{5}}$ & $\frac{1}{12\sqrt{15}}$ & $\frac{1}{6\sqrt{3}}$
			\\
			$\widehat{E}^{(3,2,\frac{1}{6})}_{(24,0)}$
			& $\widehat{\Phi}^{(\overline{3},2,-\frac{1}{6})}_{(75,0)}$
			& $0$ & $0$ & $0$ & $0$ & $0$
			& $0$ & $-\frac{\sqrt{5}}{3\sqrt{6}}$ & $-\frac{1}{3\sqrt{6}}$
			\\
			\hline
			$\widehat{\Delta}^{(3,2,\frac{1}{6})}_{(\overline{50},-2)}$
			&  $\widehat{\overline{\Delta}}^{(\overline{3},2,-\frac{1}{6})}_{(50,2)}$
			& $\frac{i}{5\sqrt{10}}$ & $-\frac{i7}{10\sqrt{15}}$
			& $0$ & $0$ & $0$ & $-\frac{1}{10\sqrt{15}}$ & $-\frac{7}{60\sqrt{15}}$ & $\frac{1}{15\sqrt{3}}$
			\\
			$\widehat{\Delta}^{(3,2,\frac{1}{6})}_{(\overline{50},-2)}$
			&  $\widehat{A}^{(\overline{3},2,-\frac{1}{6})}_{(24,0)}$
			& $0$ & $0$ & $0$ & $0$ & $-\frac{1}{5}$ & $0$ & $0$ & $0$
			\\
			$\widehat{\Delta}^{(3,2,\frac{1}{6})}_{(\overline{50},-2)}$
			&  $\widehat{\Phi}^{(\overline{3},2,-\frac{1}{6})}_{(24,0)}$
			& $0$ & $0$ & $0$ & $0$ & $-\frac{1}{30}$ & $0$ & $0$ & $0$
			\\
			$\widehat{\Delta}^{(3,2,\frac{1}{6})}_{(\overline{50},-2)}$
			&  $\widehat{\Phi}^{(\overline{3},2,-\frac{1}{6})}_{(75,0)}$
			& $0$ & $0$ & $0$ & $0$
			& $-\frac{\sqrt{2}}{15}$ & $0$ & $0$ & $0$
			\\
			\hline
			$\widehat{\Phi}^{(3,2,\frac{1}{6})}_{(24,0)}$
			& $\widehat{\Phi}^{(\overline{3},2,-\frac{1}{6})}_{(24,0)}$
			& $\frac{4\sqrt{2}}{5\sqrt{5}}$ & $-\frac{4}{15\sqrt{15}}$
			& $\frac{1}{12\sqrt{15}}$
			& $0$ & $0$ & $\frac{1}{2\sqrt{15}}$ & $-\frac{7}{18\sqrt{15}}$ & $\frac{1}{18\sqrt{3}}$
			\\
			$\widehat{\Phi}^{(3,2,\frac{1}{6})}_{(24,0)}$
			&  $\widehat{\Phi}^{(\overline{3},2,-\frac{1}{6})}_{(75,0)}$
			& $0$ & $\frac{2\sqrt{2}}{3\sqrt{15}}$
			& $-\frac{\sqrt{5}}{3\sqrt{6}}$ & $0$ & $0$ & $0$
			& $-\frac{\sqrt{5}}{9\sqrt{6}}$
			& $\frac{\sqrt{2}}{9\sqrt{3}}$
			\\
			$\widehat{\Phi}^{(3,2,\frac{1}{6})}_{(75,0)}$
			& $\widehat{\Phi}^{(\overline{3},2,-\frac{1}{6})}_{(75,0)}$
			& $-\frac{2\sqrt{2}}{5\sqrt{5}}$ & $\frac{22}{15\sqrt{15}}$
			& $\frac{11}{12\sqrt{15}}$ & $0$ & $0$
			& $-\frac{1}{\sqrt{15}}$ & $-\frac{11}{18\sqrt{15}}$ & $\frac{4}{9\sqrt{3}}$
			\\
			$\widehat{\Phi}^{(3,2,\frac{1}{6})}_{(24,0)}$
			& $\widehat{\overline{\Delta}}^{(\overline{3},2,-\frac{1}{6})}_{(50,2)}$
			& $0$ & $0$ & $0$
			& $-\frac{1}{30}$
			& $0$ & $0$ & $0$ & $0$
			\\
			$\widehat{\Phi}^{(3,2,\frac{1}{6})}_{(75,0)}$
			& $\widehat{\overline{\Delta}}^{(\overline{3},2,-\frac{1}{6})}_{(50,2)}$
			& $0$ & $0$ & $0$ & $-\frac{\sqrt{2}}{15}$ & $0$ & $0$ & $0$ & $0$
			\\
			\hline \hline
		\end{tabular}
	\end{center}
\end{table}

\end{document}